\documentstyle[psfig,lscape]{elsart}
\newcommand{\bare}[1]{\mathaccent"7017{#1}}
\def\e{\epsilon}
\def\c{{\mathrm c}}
\def\d{{\mathrm d}}
\def\p{{\mathbf p}}
\def\rot{r_{0T}}
\def\rol{r_{0L}}
\newcommand{\Li}[1]{\mbox{$\,{\mathrm Li}_2\! \left( #1 \right)$}}

\newcommand{\Dia}[3]{\raisebox{-#2mm}{\psfig {figure=#1.ps,width=#3mm}}}
\begin{document}
%
%
%
%
\begin{frontmatter}
\title{Minimal renormalization without $\e$-expansion:
Three-loop amplitude functions of the O($n$) symmetric $\phi^4$
theory in three dimensions below $T_\c$}
\author{M.~Str\"osser\thanksref{e-mail-1}},
\author{S.A.~Larin\thanksref{e-mail-2}},
\author{V.~Dohm\thanksref{e-mail-3}}
\address{Institut f\"{u}r Theoretische Physik, 
Technische Hochschule Aachen,\\
D-52056 Aachen, Germany}
\thanks[e-mail-1]{E-mail: stroesse@physik.rwth-aachen.de}
\thanks[e-mail-2]{Permanent address: 
Institute for Nuclear Research of the Russian Academy of Science, 
60$^{\mathrm th}$ October Anniversary Prospect 7-A, Moscow 117312, Russia}
\thanks[e-mail-3]{E-mail: vdohm@physik.rwth-aachen.de}
\date{\today}

\begin{abstract}
We present an analytic
three-loop calculation for thermodynamic quantities of
the O($n$) symmetric $\phi^4$ theory below $T_\c$ within the
minimal subtraction scheme at fixed dimension $d=3$.
Goldstone singularities arising at an intermediate
stage in the calculation of O($n$) symmetric quantities cancel among themselves
leaving a finite result in the limit of zero external field.
From the free energy we calculate the three-loop terms of the
amplitude functions $f_\phi$, $F_+$ and
$F_-$ of the order parameter and the specific heat above and below $T_\c$,
respectively, without using the $\e=4-d$ expansion. 
A Borel resummation for the case $n=2$ yields
resummed amplitude functions $f_\phi$ and $F_-$ that are slightly larger
than the one-loop results. Accurate knowledge of 
these functions is needed for testing the renormalization-group prediction
of critical-point universality along the $\lambda$-line of
superfluid $^4$He.
Combining the three-loop result for $F_-$ with a recent five-loop calculation
of the additive renormalization constant of the specific heat yields excellent
agreement between the calculated and measured universal amplitude ratio
$A^+/A^-$ of the specific heat of $^4$He.
In addition we use our result for $f_\phi$ to calculate the universal
combination $R_C$ of the amplitudes of the order parameter, the susceptibility
and the specific heat for $n=2$ and $n=3$. Our Borel-resummed three-loop result
for $R_C$ is significantly more accurate than the previous result
obtained from the $\e$-expansion up to $O(\e^2)$.\\[5mm]
\end{abstract}
\end{frontmatter}

\newpage
\section{Introduction}
Field-theoretic perturbative calculations of the critical behavior of O($n$)
symmetric systems with $n>1$ below $T_\c$ are known to be considerably more
complicated than those of Ising-like $(n=1)$ systems. The difficulties for
$n>1$ are due to the existence of transverse fluctuations of the order
parameter (in addition to the ordinary longitudinal fluctuations of $n=1$
systems) which, in the long-wavelength limit and at vanishing external field, 
have a vanishing restoring force.
This implies the existence of massless Goldstone modes
\cite{goldstone61,wagner66} which yield
infinite transverse and longitudinal susceptibilities \cite{pata73} and
which cause infrared (Goldstone)
singularities at intermediate stages of perturbative calculations
of all thermodynamic quantities on the coexistence curve below $T_\c$.
These complications for $n>1$ have prevented calculations of
the equation of state and of amplitude
functions below $T_\c$ to higher than two-loop 
order \cite{bervillier76,SD2,Str} , in contrast
to the case $n=1$ below $T_\c$
where accurate Borel resummation results based on five-loop
perturbation theory are available \cite{BBMN,HD,GZ-J,Lar}.

Higher-order calculations of the amplitude functions below $T_\c$ for the
case $n=2$ are of primary importance in view of the proposed theoretical
research \cite{D1} parallel to the considerable experimental effort \cite{LDID}
to test the universality prediction of the renormalization-group theory
\cite{PHA} along the $\lambda$-line of $^4$He. The amplitude functions contain
the information about universal ratios \cite{PHA}
of leading and subleading amplitudes
near criticality that have previously been used as fit parameters in the data
analysis \cite{ahlers71,lipa96} 
because of the lack of accurate predictions for these universal ratios.
The amplitude functions are also a crucial ingredient in a nonlinear
renormalization-group analysis
\cite{dohm84,D3,D2} (equivalent to a resummation of
the whole Wegner correction \cite{wegner} series)
in a wide temperature
range including non-asymptotic and non-universal effects.

In the present paper we perform the next necessary step towards the goal of an
accurate determination of amplitude functions by presenting the analytic
results of a three-loop calculation within the O($n$) symmetric $\phi^4$
theory above and 
below $T_\c$ for general $n$. Specifically, from the free energy we 
shall derive the amplitude function $f_\phi$ of the order parameter
and the amplitude functions $F_\pm$ of the specific
heat above and below $T_\c$ in the limit of vanishing external field.
The conceptual framework of our calculation is the
minimally renormalized massive $\phi^4$ field theory at fixed dimension 
$d=3$ \cite{SD2,D3,SD1}, without using the $\e=4-d$ expansion,
involving an appropriately defined pseudo-correlation legth $\xi_-$
\cite{SD2,Str} that is applicable to both $n=1$ and $n>1$ below $T_\c$.
Our $d=3$ 
approach differs from Parisi's \cite{parisi} original suggestion of the $d=3$
field theory and from the $d=3$ field theory of subsequent work
\cite{BBMN,GZ-J,BNGM,BB,muenster}
where renormalization conditions rather than the minimal
renormalization scheme were used. Our combination of the minimal subtraction
scheme with the field theory at fixed $2<d<4$ has several advantages, such as
the independence of renormalization constants on whether $T>T_\c$ or $T<T_\c$
which implies a natural decomposition of correlation functions into amplitude
functions and exponential parts where the structure of the latter is
independent of whether $T>T_\c$ or $T<T_\c$. The same exponential parts can
then be used, without modification, in extensions of the theory to critical
dynamics \cite{dohm91} or to critical phenomena in confined systems above
and below $T_\c$ \cite{esser}.

Our approach has
been successfully employed recently \cite{Str} in deriving various amplitude
functions, including $f_\phi$ and $F_\pm$,
for general $n$ below $T_\c$ in two-loop order.
In three-loop order, the technical difficulties related to the removal of 
ultraviolet divergences in three dimensions and to the treatment of spurious
infrared (Goldstone) divergences are substantially greater.
The concept of the minimally renormalized $d=3$ field theory turns out to
constitute an appropriate framework for coping with 
these difficulties. This requires to further
develop new integration techniques based on recent advances by Rajantie
\cite{Raj}
in the analytic evaluation of three-loop integrals in three dimensions.
In particular we succeed in calculating two
``Mercedes'' diagrams with two different
(longitudinal and transverse) masses that were not considered previously.

The perturbation series of the $\phi^4$ theory are known to be divergent and
to require resummations \cite{zinn81} in order to yield reliable results.
The three-loop terms derived in the
present paper make possible to perform Borel resummations that lead to results
with reasonably small error bars. In a related paper \cite{Lar}
this was shown for $F_-$. In the present paper we demonstrate this for the
case of $f_\phi$.
For the example $n=2$ the Borel resummed
three-loop amplitude functions $F_-$ and $f_\phi$
of the specific heat below $T_\c$ and of the
order parameter turn out to be slightly larger than the one-loop results.
Application of the Borel resummed amplitude function $F_-$
to the universal amplitude ratio $A^+/A^-$ of the asymptotic
specific heat yields the theoretical
prediction \cite{Lar} $A^+/A^-=1.056\pm0.004$ which
is in excellent agreement with the high-precision experimental result 
\cite{lipa96} $A^+/A^-=1.054\pm0.001$ for $^4$He near the superfluid
transition obtained from a recent experiment in space.
Furthermore we use our result for $f_\phi$ to calculate the universal
combination $R_C$ \cite{PHA} of the amplitudes of the order parameter, the
susceptibility and the specific heat for $n=2$ and $n=3$. Our Borel
resummed three-loop result for $R_C$ is significantly more accurate than 
the previous result obtained from the $\e$-expansion up to $O(\e^2)$
\cite{AH}.

\section{Bare Helmholtz free energy}
\label{sec:bare}
\begin{figure}
\hspace{-21mm}\psfig{figure=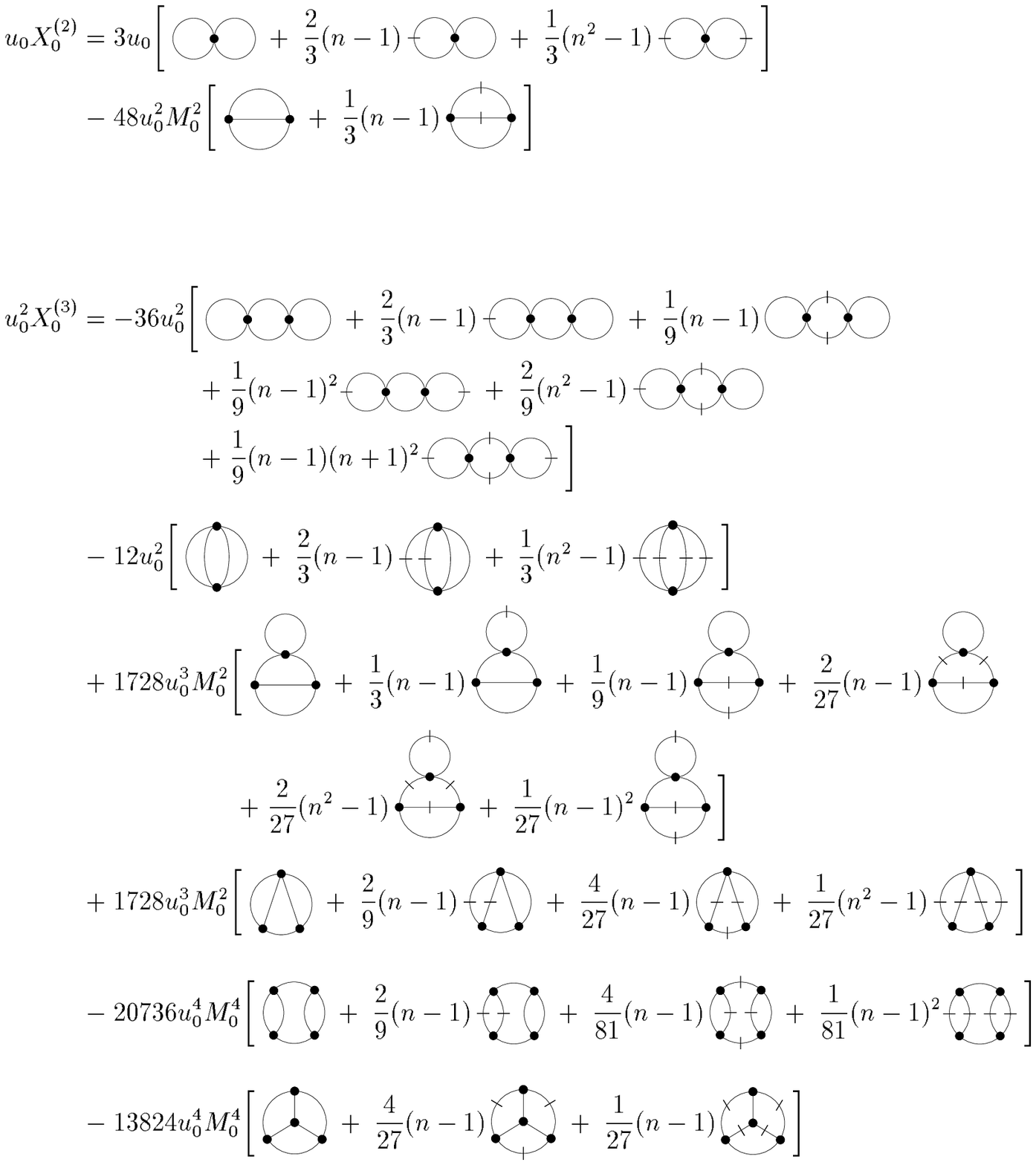,width=16cm}

\vspace{-6cm}

\caption{
Two- and three-loop vacuum diagrams determining the Helmholtz free energy
$\Gamma_0(r_0,u_0,M_0,\Lambda)$, Eq.~(\protect\ref{eq:Gamma_3loop}). 
The lines denote longitudinal and transverse propagators,
$G_L(p)=$ \protect\rule[2.2pt]{30pt}{0.1mm}\, $\equiv (\bar{r}_{0L}+p^2)^{-1}$
and
$G_T(p)=$ \protect\rule[2.2pt]{30pt}{0.1mm}\hspace{-19pt}
\protect\rule{0.1mm}{5pt}\hspace{17pt} $\equiv (\bar{r}_{0T}+p^2)^{-1}$.
The integral expressions are given in Appendix \ref{app:diagrams}. The analytic
results for the last two ``Mercedes'' diagrams at $d=3$ are given in
Eqs.~(\ref{eq:Merc1}) and (\ref{eq:Merc2}).
}
\label{fig:vacuum}
\end{figure}

Throughout this paper we shall use the notation and definitions of 
Ref.~\cite{Str}. We start from the O($n$) symmetric $\phi^4$ model with the
Landau-Ginzburg-Wilson functional
\begin{equation}
\label{eq:LGW}
{\cal H}\{\vec\phi_0({\bf x})\} = \int_V \mbox{d}^{d}x
\left(\frac{1}{2}r_0\phi_0^2 + \frac{1}{2}\sum_{i}
(\nabla\phi_{0i})^2
+u_0(\phi_0^2)^2 -\vec{h}_0\cdot\vec\phi_0\right) 
\end{equation}
for the $n$-component field 
$\vec\phi_0({\bf x})=(\phi_{01}({\bf x}),\ldots,\phi_{0n}({\bf x}))$
in the presence of the homogeneous external field
$\vec{h}_0=(h_0,0,\ldots,0)$.
The spatial 
variations of 
$\vec\phi_0({\bf x})$ are restricted to wavenumbers less than some cutoff 
$\Lambda$.
We are interested in the bulk Helmholtz free energy
$\Gamma_0(r_0,u_0,M_0,\Lambda)$ per unit volume
with $M_0\equiv\langle\phi_{01}\rangle$. $\Gamma_0$ is obtained from the
negative sum of all one-particle irreducible vacuum diagrams.
The structure of the analytic expression is (apart from 
an unimportant additive constant)
\begin{eqnarray}
\label{eq:Gamma_3loop}
\Gamma_0(r_0,u_0,M_0,\Lambda) &=&
   \frac{1}{2} r_0M_0^2 +u_0M_0^4
   +\frac{1}{2}\int_\p^\Lambda \ln \left( \bar{r}_{0L}+p^2 \right) \nonumber\\
&& \mbox{}
   +\frac{1}{2}(n-1)\int_\p^\Lambda \ln \left( \bar{r}_{0T}+p^2 \right) 
   + u_0X_0^{(2)}(r_0,u_0,M_0,\Lambda) 
   \nonumber\\
&& \mbox{} +u_0^2X_0^{(3)}(r_0,u_0,M_0,\Lambda) +O(u_0^3)
\label{eq:rol_bar}
\end{eqnarray}
where $\int_\p^\Lambda\equiv (2\pi)^{-d}\int^\Lambda {\mathrm d}^dp$
means integration up to $|\p|<\Lambda$. The terms
$u_0X_0^{(2)}(r_0,u_0,M_0,\Lambda)$ and $u_0^2X_0^{(3)}(r_0,u_0,M_0,\Lambda)$ 
represent
the two- and three-loop contributions shown in Fig.~\ref{fig:vacuum},
with longitudinal and transverse propagators $G_L(p)=(\bar{r}_{0L}+p^2)^{-1}$ 
and $G_T(p)=(\bar{r}_{0T}+p^2)^{-1}$ where
\begin{equation}
\label{eq:bar(rol)}
\bar{r}_{0L} = r_0+12u_0M_0^2\,, \quad \bar{r}_{0T} = r_0+4u_0M_0^2
\end{equation}
(compare Eq.~(27) and Fig.~3 of Ref.~\cite{Str} at $k=0$).
For the integral expressions of $X_0^{(3)}$ see Appendix \ref{app:diagrams}.

We consider the limit $\Lambda\to\infty$ (although neglecting cutoff effects
may not be justified in certain cases \cite{dohm93,bagnuls94,anisimov95}).
The right-hand side of Eq.~(\ref{eq:Gamma_3loop}) is ultraviolet divergent
for $\Lambda\to\infty$ at any $d>2$. To absorb these divergences requires
two different steps: a shift of the temperature variable $r_0$ (mass shift)
and a subtraction from the free energy.
The first step is to turn to the shifted variable $r_0-r_{0c}$ where
$r_{0\c}(u_0,\Lambda)$ is the critical value of $r_0$ \cite{SD1}. Substituting
$r_0=r_0-r_{0\c}+r_{0\c}(u_0,\Lambda)$ into $\Gamma_0$ yields the function
\begin{equation}
\label{eq:Gamma_hut}
\hat{\Gamma}_0(r_0-r_{0\c},u_0,M_0,\Lambda)
= \Gamma_0(r_0-r_{0\c}+r_{0\c}(u_0,\Lambda),u_0,M_0,\Lambda) \,.
\end{equation}
(Because of the non-analytic $u_0$ dependence of $r_{0\c}$, the function
$\hat{\Gamma}_0$ does not have an expansion in integer powers of $u_0$.)
This function is still divergent for $d>2$ in the limit
$\Lambda\to\infty$ at fixed
$r_0-r_{0\c}$, $u_0$, $M_0$. We have verified up to three-loop order that
these ultraviolet divergences have a regular
dependence on $r_0-r_{0\c}$
(up to linear order
in $r_0-r_{0\c}$ for $d<4$ and up to quadratic order in $r_0-r_{0\c}$
for $d=4$).
Therefore it suffices to perform additional overall subtractions from
$\hat{\Gamma}_0$ that are regular in $r_0-r_{0\c}$.
They do not affect the singular
part of the temperature dependence of the free energy.
The resulting singular part of the bare free energy is finite in the limit
$\Lambda\to\infty$ for $2<d<4$ at fixed $r_0-r_{0\c}$.

Here we use the prescriptions of dimensional regularization at 
$\Lambda=\infty$ and employ the critical parameter $r_{0\c}$ in the form
\cite{SD1,symanzik}
\begin{equation}
\label{r0c}
r_{0\c}(u_0,\e) = u_0^{2/\e} S(\e)
\end{equation}
where the function $S(\e)$ is finite for $\e>0$ except for poles at
$d_l=4-2/l$, $l=2$, 3, \ldots
In our version of the $d=3$ theory it is sufficient to work near $d=3$ and,
instead of $r_0-r_{0\c}$, to
use a shifted variable $r_0'$ as defined by
$r_0=r_0'+\delta r_0(u_0,\e)$ with a simplified $\delta r_0(u_0,\e)$ 
that contains
only the $d=3$ pole of $r_{0\c}$ (and not the poles at $d_l>3$ with $l\geq3$).
Thus we use \cite{Str,BBMN,HD}
\begin{equation}
\label{eq:r0'}
\delta r_0(u_0,\e) = \frac{1}{\pi^2}(n+2)\frac{u_0^{2/\epsilon}}{\epsilon -1}
           + C(n)\, u_0^{2/\epsilon}
\end{equation}
with $\e=4-d$ and the finite constant
\begin{equation}
\label{eq:C(n)}
C(n) = \frac{n+2}{\pi^2} \left[ 1 - C_{\mathrm Euler}
+ \ln \frac{4\pi}{9} -2\ln24 \right]
\end{equation}
(the final results for the amplitude function do not depend
on the choice of $C(n)$ \cite{Str,SD1}).
Correspondingly, instead of $\bar{r}_{0L}$
and $\bar{r}_{0T}$, we use the longitudinal and transverse parameters
\begin{equation}
\label{eq:rol}
\rol = r_0'+12u_0M_0^2\,, \quad \rot = r_0'+4u_0M_0^2\,.
\end{equation}
In the limit $d\to3$ we then obtain the singular part of the bare
Helmholtz free energy as a finite
function $\bare{\Gamma}(r_0',u_0,M_0)$ whose two-loop expression has been
derived recently (see Eq.~(31) of Ref.~\cite{Str}).

We have been able to perform an analytic calculation
of the three-loop contributions to $\bare{\Gamma}(r_0',u_0,M_0)$ near the
coexistence curve below $T_\c$ and for $M_0^2=0$ above $T_\c$.
The details of this calculation will be presented elsewhere.
Some comments are given in Appendix \ref{app:diagrams}.
The resulting bare perturbation expression for the Helmholtz
free energy at $r_0'\not=0$ reads up to three-loop order at $d=3$
\begin{eqnarray}
\label{eq:free_energy}
\bare{\Gamma}(r_0',u_0,M_0) &=& \frac{1}{2}r_0' M_0^2+u_0M_0^4
                + \sum_{b=1}^3\, \sum_{l=0}^{b-1}\, \sum_{k=0}^1\,
                (-1)^k\, 2^{-l-k} F_{blk}(\bar{w},n) \nonumber\\
   && \mbox{}   \cdot\, (24u_0)^{3-l} (M_0^2)^l\, \left[
                \frac{\rol}{(24u_0)^2}\right]^{\textstyle\frac{4-b-2l}{2}}
                \left(\ln \frac{\rol}{(24u_0)^2}\right)^{\! k} 
\end{eqnarray}
where
\begin{equation}
\label{eq:w}
\bar{w}(r_0',u_0,M_0)=\rot/\rol
\end{equation}
is a non-perturbative parameter. Above $T_\c$ its largest possible value is 
{$\bar{w}(r_0',u_0,0)=1$} (when
$r_0'>0$ and $M_0^2=0$ corresponding to $T>T_\c$ at $h_0=0$).
The smallest possible value of $\bar{w}$ is attained on the coexistence curve
well below $T_\c$ where, at given $r_0'<0$ and to lowest order in $u_0$,
$\bar{w}(r_0',u_0,M_0(r_0',u_0))=3\pi^{-1}u_0(-2r_0')^{-1/2}+O(u_0^2)$
as follows from Eq.~(\ref{eq:M0(r0')}) in Sect.~\ref{sec:order}.
The coefficients $F_{blk}(\bar{w},n)$ depend on $\bar{w}$ and $n$
and determine the contributions in $b$-loop order.
The one- and two-loop coefficients are \cite{Str}
\begin{eqnarray}
\label{eq:F100}
F_{100}(\bar{w},n) 
&=& -\frac{1}{12\pi}\, \left[ 1+(n-1)\bar{w}^{3/2} \right] \,,\\ 
\label{eq:F200}
F_{200}(\bar{w},n) 
 &=& \frac{1}{384\pi^2}\, \left[ 3 +2(n-1)\bar{w}^{1/2} +(n^2-1)\bar{w} 
            \right] \,,\\ 
\label{eq:F210}
F_{210}(\bar{w},n) 
 &=& \frac{1}{288\pi^2}\, (n-1)\ln\frac{1+2\bar{w}^{1/2}}{3} \,,\\
\label{eq:F211}
F_{211}(n) 
 &=& -\frac{1}{288\pi^2}\, (n+2) \,.
\end{eqnarray}
The new three-loop coefficients $F_{300}$ and $F_{301}$ read in analytic form
\begin{eqnarray}
\label{eq:F300}
F_{300}(\bar{w},n) 
 &=& \frac{1}{18432\pi^3}\, \Bigg\{ 15 +24\ln\frac{3}{4} 
        - (n-1) \Bigg[ \bar{w}^{-1/2} +2n -6 \nonumber\\       
&& \mbox{}
        +8\ln\frac{2+2\bar{w}^{1/2}}{3} 
        +\bar{w}^{1/2} \Bigg( n^2-6n-9
        +4(n+1) \ln\frac{16\bar{w}}{9}
        \nonumber\\
&& \left.\mbox{} 
        +8\ln \frac{2+2\bar{w}^{1/2}}{3} \Bigg) 
        +\bar{w}(n-1) \Bigg] \right\} \,,\\
\label{eq:F301}
F_{301}(\bar{w},n)
 &=& \frac{1}{2304\pi^3}\, \left\{ 3 +(n-1)\left[ 1 +(n+2)\bar{w}^{1/2}
            \right] \right\} \,.
\end{eqnarray}
In calculating the coefficients $F_{310}$ and $F_{320}$ below $T_\c$ we have
confined ourselves to the vicinity of the coexistence curve where an expansion
with respect to $\bar{w}$ is justified. 
For the purpose of calculating the order parameter this expansion must include
the terms of linear order in $\bar{w}$ whereas for the calculation of the
specific heat all terms can be neglected that vanish in the limit
$\bar{w}\to0$. The latter statement applies to $F_{300}$ and $F_{301}$ as well.
The result for the coefficients $F_{310}$ and $F_{320}$ is
\begin{eqnarray}
\label{eq:F310-}
F_{310}(\bar{w},n) 
 &=& \frac{1}{27648\pi^3}\, \Bigg\{ 9\pi^2 -18 
            +108\Li{-\frac{1}{3}} -(n-1) \Bigg[ 4\bar{w}^{-1/2} \nonumber\\
&& \mbox{}  +4n+2 -(n+2)\pi^2 -12\Li{\frac{1}{3}}
            -32\ln2 -6\,(\ln3)^2 
            \nonumber\\
&& \mbox{}  + \bar{w}^{1/2} \Big( 10n+32 
            -16\,(2n+3)\ln2 +48\ln3 -8(n+1)\ln\bar{w} \Big)
             \nonumber\\
&& \mbox{}  + \frac{1}{3}\bar{w} \Big( 84n-100-128\ln2 \Big)
            +O(\bar{w}^{3/2},\bar{w}^{3/2}\ln\bar{w})
            \Bigg] \Bigg\} \,,\\
\label{eq:F320-}
F_{320}(\bar{w},n) 
 &=& \frac{1}{165888\pi^3}\, \Bigg\{ 432\ln\frac{4}{3}
       -324\Li{-\frac{1}{3}} -432c_1 -27\pi^2 \nonumber\\
&&\mbox{}  - (n-1) \Bigg[ 16\bar{w}^{-1/2} + \frac{3n+14}{3} \pi^2 
       +18(\ln3)^2 +36\Li{\frac{1}{3}} \nonumber\\
&&\mbox{}  +16\!\left( c_2 +4\Li{-2} -2\Li{-\frac{1}{2}}
       +\!\!\!\;\left[6\ln3-\ln2-\frac{13}{3}\right]\ln2\right) \nonumber\\
&&\mbox{} -\frac{128}{3}
      +16\bar{w}^{1/2} \Big( 7-n +(n+1)\ln\,(16\bar{w}) +2\ln2-6\ln3 
      \Big) \nonumber\\
&&\mbox{}
      +4\bar{w} \Bigg( 4c_2 -12n -\frac{224}{5} +\pi^2
      +6\left[ 6\ln3 -\ln2 -\frac{16}{15} \,\right]\ln2 \nonumber\\
&&\mbox{}
      +12 \left[ 2\Li{-2} -\Li{-\frac{1}{2}} \right] \Bigg)
      +O(\bar{w}^{3/2},\bar{w}^{3/2}\ln\bar{w}) \Bigg] \Bigg\} \,.
\end{eqnarray}
Here $\Li x \equiv \int_x^0 t^{-1} \ln {(1-t)} \,\d t$ is the dilogarithmic
function. The constants $c_1$ and $c_2$ are given by
\begin{eqnarray}
\label{eq:c1}
c_1 &=& \int_0^1\frac{\d x}{\sqrt{6-2x^2}} \left[ \ln 
\frac{3}{4} +\ln \frac{3+x}{2+x} +\frac{x}{2+x} \left( \ln \frac{3+x}{3}
+\frac{x}{2-x} \ln \frac{2+x}{4} \right) \right] \nonumber\\
&=& 0.0217376333
\end{eqnarray}
and
\begin{equation}
\label{eq:c2}
c_2 = \frac{\pi^2}{4\sqrt{2}} +\sqrt{2} \int_0^1 \frac{\d x}{\sqrt{1+x^2}}
\left[ \ln \frac{x}{1+x} +\frac{\ln(1+x)}{x} \right]
=0.973771427 \,.
\end{equation}
Above $T_\c$, at $M_0^2=0$, the coefficients in 
Eqs.~(\ref{eq:F100})--(\ref{eq:F301}) are taken at $\bar{w}=1$ and
the coefficients $F_{310}$ and $F_{320}$ do not contribute.
All other coefficients $F_{3lk}$ vanish.
Eqs.~(\ref{eq:free_energy})--(\ref{eq:c2}) are
the main result of our paper. They provide the basis
for deriving the analytic form of the
amplitude functions of the order parameter and the specific
heat for general $n$.
For the case $n=1$, Eqs.~(\ref{eq:free_energy})--(\ref{eq:c2})
reduce to the three-loop part of Eq.~(3.18) of Bagnuls et al.\ \cite{BBMN} and
of Eq.~(3.3) of Halfkann and Dohm \cite{HD} and agree with the numerical 
values of their coefficients
$F_{blk}$ for $b=1$, $b=2$ and $b=3$. The latter agree also with those in Eq.\
(A1.1) of Guida and Zinn-Justin \cite{GZ-J} for $n=1$.

Eq.~(\ref{eq:free_energy}) 
contains logarithms of the coupling $u_0$ as expected
because of the non-analytic $u_0$ dependence of $\delta r_0$, 
Eq.~(\ref{eq:r0'}).
Furthermore, the terms proportional to $n-1$ depend non-analytically on $\rot$
through $\bar{w}^{-1/2}$, $\bar{w}^{1/2}$ and $\bar{w}^{1/2}\ln\bar{w}$ 
(and higher orders in the neglected terms).
These non-analyticities will lead to perturbative terms in the derivatives of 
the free energy (with respect to $M_0$)
that diverge when the coexistence curve is approached ($T<T_\c$,
$h_0\to0$). This is again the effect of the
Goldstone modes that was found previously in two-loop order \cite{Str}. For
O($n$) symmetric quantities, however,
these divergences should cancel among themselves \cite{elitzur83}.
This is indeed the case, at least up to three-loop order,
for the square of the order parameter, the Gibbs free energy
and for the specific heat as we shall see below.
\section{Bare order parameter}
\label{sec:order}
The order parameter $M_0(r_0',u_0,h_0)$ 
is determined by inverting the equation of state,
\begin{equation}
\label{eq:eos}
h_0(r_0',u_0,M_0)=
\frac{\partial}{\partial M_0}\bare{\Gamma}(r_0',u_0,M_0)\,.
\end{equation}
This inversion should be performed iteratively at finite $h_0\not=0$.
Several perturbative
terms of $\partial \bare{\Gamma} / \partial M_0$ exhibit Goldstone
singularities for $h_0\to0$ below $T_\c$ which arise from the three-loop terms 
$\sim\bar{w}^{-1/2}$, $\bar{w}^{1/2}$ and $\bar{w}^{1/2}\ln\bar{w}$
(after an expansion of $\bar{w}$ dependent terms
with respect to $u_0$ at $h_0\not=0$),
in addition to the known \cite{Str} two-loop singularities,
as well as from the expansion of one- and two-loop terms 
(with respect to $u_0$ contained in $M_0^2(r_0',u_0,h_0)$ at finite $h_0$). 
We have verified that all Goldstone divergences cancel among
themselves which leads to a finite result for $M_0^2$ as a function of $r_0'$
and $u_0$ in the limit $h_0\to0$ at $d=3$. This function reads for $r_0'<0$
\begin{eqnarray}
\label{eq:M0(r0')}
M_0^2 &=& \frac{1}{8u_0}(-2r_0')+\frac{3}{4\pi}(-2r_0')^{1/2} \nonumber\\
      && \mbox{} 
          +\frac{u_0}{8\pi^2} \left[ 10 -n +4(n-1)\ln 3 -2(n+2)
          \ln \frac{-2r_0'}{(24u_0)^2} \right]
          \nonumber\\
      &&  \mbox{}
          +\frac{u_0^2(-2r_0')^{-1/2}}{1920\pi^3} \,\Bigg\{
          -2736n -5904 -6480c_1 +240(n-1)c_2 \nonumber\\
      &&  \mbox{}
          -(75n^2-5n+875)\pi^2 -1260 \left[ (n-1)\Li{\frac{1}{3}} 
          +9\Li{-\frac{1}{3}} \right] \nonumber\\
      &&  \mbox{}
          +960(n-1) \left[ 2\Li{-2} -\Li{-\frac{1}{2}} \right]
          -630(n-1) (\ln 3)^2 
          \nonumber\\
      &&  \mbox{}
          -48\ln2 \Big[
          10(n-1)\ln2 -60(n-1)\ln3 +111n-561 \Big] \nonumber\\
      &&  \mbox{}
          +240(12n-57)\ln3 
          -1440(n+2) \ln \frac{-2r_0'}{(24u_0)^2} \,\Bigg\} \nonumber\\
      &&  \mbox{}
          +O(u_0^3,u_0^3\ln u_0) \,.
\end{eqnarray}
It contains the expected \cite{Str}
logarithmic $u_0$ dependence $\sim u_0^2\ln u_0$ at three-loop order.
The logarithmic terms in $u_0$ 
can be absorbed by employing the pseudo-correlation
length \cite{SD2,Str} $\xi_-$ as a temperature variable below $T_\c$ instead
of $r_0'<0$. The relation between $r_0'$ and $\xi_-$ is determined by the
two-point vertex function and
is given up to three-loop order by (see Appendix \ref{app:correlation})
\begin{eqnarray}
\label{eq:r0'(xi-)}
-2r_0'  = \xi_-^{-2} \Bigg\{ && 1 + \frac{n+2}{\pi} u_0\xi_- 
          -\frac{n+2}{\pi^2} (u_0\xi_-)^2 \left[
          \frac{1385}{108} +4 \ln(24u_0\xi_-) \right] \nonumber \\
&& \mbox{} + \frac{n+2}{108\pi^3} (u_0\xi_-)^3 \Bigg[
           1314n +13047 +576(n+8) \Li{-\frac{1}{3}}
           \nonumber\\
&& \mbox{} +48(n+8) \pi^2 +8\, (43n+182) \ln \frac{3}{4} \,\Bigg]
           + O(u_0^4\xi_-^4) \,\Bigg\} \,.
\end{eqnarray}
This leads to the bare
perturbative expression for the square of the order parameter 
$M_0^2(\xi_-,u_0,3)$ that is finite in three dimensions and is free of
logarithms of $u_0$,
\begin{equation}
\label{eq:M0(xi-)}
M_0^2(\xi_-,u_0,3) = \xi_-^{-1} \left\{ \sum_{m=0}^3 a_{\varphi m} 
                     (u_0\xi_-)^{m-1}
                     +O(u_0^3\xi_-^3) \right\}
\end{equation}
with the coefficients up to two-loop order \cite{Str},
\begin{eqnarray}
\label{eq:a_phi0}
a_{\varphi 0} &=& \frac{1}{8} \,,\\
\label{eq:a_phi1}
a_{\varphi 1} &=& \frac{1}{8\pi}(n+8) \,,\\
\label{eq:a_phi2}
a_{\varphi 2} &=& \frac{1}{2\pi^2}(n-1)\ln 3 -\frac{1}{864\pi^2}(1169n+1042) 
                  \,,
\end{eqnarray}
and the new three-loop coefficient
\begin{eqnarray}
\label{eq:a_phi3}
a_{\varphi 3} &=& \frac{1}{17280\pi^3} \Bigg\{
            36\,(565n^2 +5056n +7744) -58320c_1 +2160(n-1)c_2 
            \nonumber\\
&&  \mbox{} 
            +15\pi^2(19n^2 +643n +499) 
            +180\, (64n^2 +640n +457) \Li{-\frac{1}{3}}
            \nonumber\\
&&  \mbox{}  
            -11340(n-1) \Li{\frac{1}{3}}
            +8640(n-1) \left[ 2\Li{-2} -\Li{-\frac{1}{2}} \right]
            \nonumber\\
&&  \mbox{}
            +80\,(86n^2+860n-811)\ln3 -16\,(860n^2+8357n-7867)\ln2 \nonumber\\
&&  \mbox{}
            -5670(n-1) (\ln3)^2 +4320(n-1) \Big( 6\ln3 -\ln2 \Big) \ln2
            \,\Bigg\} \,.
\end{eqnarray}
For $n=1$, Eqs.~(\ref{eq:M0(xi-)})--(\ref{eq:a_phi3}) agree with Eq.~(3.13) 
and the numerical values in Table 2 of Halfkann and Dohm \cite{HD}.

\section{Bare Gibbs free energy}
\label{sec:Gibbs}
The bare Gibbs free energy $\bare{\cal{F}}$ is determined by the bare
Helmholtz free energy and the order parameter as
\begin{equation}
\label{eq:Gibbs}
\bare{\cal F}(r_0',u_0,h_0)=
\bare{\Gamma}(r_0',u_0,M_0(r_0',u_0,h_0)) -h_0M_0(r_0',u_0,h_0) \,.
\end{equation}
Above $T_\c$ we obtain the Gibbs free energy $\bare{\cal F}_+(r_0',u_0)$
at $h_0=0$ by substituting $M_0=0$ and $\bar{w}=1$ into $\bare{\Gamma}$.
From Eqs.~(\ref{eq:free_energy})--(\ref{eq:F301}) we find at $d=3$
\begin{eqnarray}
\label{eq:Gamma+(r0')}
\bare{\cal F}_+(r_0',u_0) &=& \bare{\Gamma}(r_0',u_0,0) \nonumber\\
&=& -\frac{n}{12\pi} r_0'^{\,3/2}
                              -\frac{n(n+2)}{(4\pi)^2} u_0 r_0' 
                              -\frac{2n(n+2)}{(4\pi)^3} u_0^2 r_0'^{\,1/2}
                              \Bigg[ n-6 \nonumber\\
&& \mbox{}                    -8\ln\frac{3}{4} +4\ln \frac{r_0'}{(24u_0)^2}
                              \Bigg] + O(u_0^3,u_0^3\ln u_0)
\end{eqnarray}
for $r_0'>0$. Below $T_\c$ we make a perturbative expansion of $\bare{\cal F}$,
Eq.~(\ref{eq:Gibbs}), in $u_0$ at $h_0\not=0$ and take the limit
\begin{equation}
\label{eq:Gamma-}
\bare{\cal F}_-(r_0',u_0) \equiv \lim_{h_0\to0} 
\bare{\cal F}(r_0',u_0,h_0)\,, \quad r_0'<0\,.
\end{equation}
The result is at $d=3$
\begin{eqnarray}
\label{eq:Gamma-(r0')}
\bare{\cal F}_-(r_0',u_0) &=& 
             -\frac{(-2r_0')^2}{64u_0} -\frac{(-2r_0')^{3/2}}{12\pi}
             \nonumber\\
  && \mbox{} -\frac{u_0(-2r_0')}{16\pi^2} 
             \left[ 6 +2(n-1)\ln 3
             -(n+2)\ln \frac{-2r_0'}{(24u_0)^2} \,\right]
             \nonumber\\
  && \mbox{} +\frac{u_0^2(-2r_0')^{1/2}}{384\pi^3} \Bigg[
             16\,(11n+7) -1296c_1  -48(n-1)c_2 
             \nonumber\\
  && \mbox{} +(9n^2+n+17)\pi^2 +36(n-1) \Li{\frac{1}{3}}
             +324 \Li{-\frac{1}{3}} 
             \nonumber\\
  && \mbox{}
             +96(n-1) \left[ \Li{-\frac{1}{2}} -2\Li{-2} \right]
             +18(n-1) (\ln3)^2 \nonumber\\
  && \mbox{}
             +48(n-1) \Big( \ln2 -6\ln3 \Big) \ln2
             -48\,(4n+17)\ln3 \nonumber\\
  && \mbox{} +16\, (31n+95)\ln2 
             +96(n+2)\ln \frac{-2r_0'}{(24u_0)^2} \,\Bigg] \nonumber\\
  && \mbox{} +O(u_0^3,u_0^3\ln u_0) \,.
\end{eqnarray}
As expected, all Goldstone divergences cancel among themselves in this limit.
The terms of
$O(\bar{w}^{1/2})$, $O(\bar{w}^{1/2}\ln\bar{w})$ and $O(\bar{w})$
in Eqs.~(\ref{eq:F300})--(\ref{eq:F320-}) do not contribute to the free energy
$\bare{\cal F}_-(r_0',u_0)$ and to the specific heat $\partial^2
\bare{\cal F}_-(r_0',u_0)/(\partial r_0')^2$
on the coexistence curve because these terms vanish in the limit $h_0\to0$
$(\bar{w}\to0)$. Therefore, for the purpose of an application to the specific
heat in Ref.~\cite{Lar}, it was sufficient to evaluate most of the diagrams
in Fig.\ 1 directly at $\rot=0$ $(\bar{w}=0)$, see also App.\ A.
This simplification, however, is not applicable to the calculation of the 
order parameter performed in the present paper.

We note that the function $\bare{\cal F}_-(r_0',u_0)$ can also be considered
as the Helmholtz free energy $\bare{\Gamma}(r_0',u_0,M_0)$ below $T_\c$ in the
limit where $M_0$ approaches the spontaneous value $M_0(r_0',u_0)$,
Eq.~(\ref{eq:M0(r0')}), of the
order parameter on the coexistence curve. Thus the Helmholtz free energy
remains finite in this limit, as expected \cite{elitzur83}.
In order to express $\bare{\cal F}_\pm$ in terms of the correlation lengths
$\xi_\pm$ we use (see Appendix \ref{app:correlation})
\begin{eqnarray}
\label{eq:r0'(xi+)}
r_0'  =    \xi_+^{-2} \Bigg\{ && 1 + \frac{n+2}{\pi}u_0\xi_+ 
           + \frac{n+2}{\pi^2} (u_0\xi_+)^2 \left[
           \frac{1}{27} +2 \ln(24u_0\xi_+) \right] \nonumber\\
&& \mbox{} + \frac{n+2}{54\pi^3} (u_0\xi_+)^3 \Bigg[
           3\,(3n+22) -144(n+8) \Li{-\frac{1}{3}} \nonumber\\
&& \mbox{} -12(n+8)\pi^2  -2\, (43n+182)\ln \frac{3}{4} \Bigg]
           + O(u_0^4\xi_+^4) \,\Bigg\}
\end{eqnarray}
for $r_0'>0$ and $r_0'(\xi_-,u_0)$ as given in Eq.~(\ref{eq:r0'(xi-)}) for
$r_0'<0$.
(We recall that, for $d=3$, the right-hand side of Eq.~(\ref{eq:r0'(xi+)})
contains a logarithmic $u_0$ dependence only at
$O(u_0^2)$, as can be seen in Eq.~(4.5) of Ref.~\cite{SD1} for $\e=1$.
The same comment applies to Eq.~(\ref{eq:r0'(xi-)}).)
Substituting Eqs.~(\ref{eq:r0'(xi+)}) and (\ref{eq:M0(r0')}) 
into Eqs.~(\ref{eq:Gamma+(r0')}) and (\ref{eq:Gamma-(r0')})
yields the Gibbs free energy up to three-loop order
\begin{eqnarray}
\label{eq:Gamma(xi+)}
\bare{\cal F}_+(\xi_+,u_0) &=& \xi_+^{-3} \Bigg\{
                 -\frac{n}{12\pi} -\frac{3n(n+2)}{16\pi^2}\,(u_0\xi_+)
                 -\frac{n(n+2)}{8\pi^3} \left[ \frac{1}{27} +n \right.
                 \nonumber\\
 &&\mbox{}       \phantom{\xi_+^{-3} \Bigg\{}
                 \left. +\ln\frac{16}{9} \right] (u_0\xi_+)^2
                 +O\Big(u_0^3\xi_+^3,u_0^3\xi_+^3\ln(u_0\xi_+)\Big) \Bigg\}
\end{eqnarray}
and
\begin{eqnarray}
\label{eq:Gamma(xi-)}
\bare{\cal F}_-(\xi_-,u_0) &=& \xi_-^{-3} \left\{
                               \sum_{m=0}^3 a^{(\Gamma)}_{-m}(u_0\xi_-)^{m-1} 
                            +O\Big(u_0^3\xi_-^3,u_0^3\xi_-^3\ln(u_0\xi_-)\Big)
                               \right\}
\end{eqnarray}
with the coefficients up to two-loop order
\begin{eqnarray}
\label{eq:a^Gamma-0}
a_{-0}^{(\Gamma)} &=& -\frac{1}{64}\,, \\
\label{eq:a^Gamma-1}
a_{-1}^{(\Gamma)} &=& -\frac{1}{96\pi}\, (3n+14)\,, \\
\label{eq:a^Gamma-2}
a_{-2}^{(\Gamma)} &=& -\frac{1}{3456\pi^2}\, \Big[54n^2-737n-394 +432(n-1)\ln3
                      \Big]
\end{eqnarray}
and the new three-loop coefficient
\begin{eqnarray}
\label{eq:a^Gamma-3}
a_{-3}^{(\Gamma)} &=& 
           \frac{1}{3456\pi^3} \Bigg[ 179n^2-3875n-10626 
           +(33n^2-471n-615)\pi^2 \nonumber\\
&& \mbox{} -11664c_1 -432(n-1)c_2 -12\,(48n^2+480n+525)\Li{-\frac{1}{3}}
           \nonumber\\
&& \mbox{} +540(n-1)\Li{-\frac{1}{2}} -1728(n-1)\Li{-2} \nonumber\\
&& \mbox{} +54(n-1) \Big[ 5\ln2-42\ln3 \Big] \ln2 +16\,(43n^2+547n+1219)\ln2
           \nonumber\\
&& \mbox{} -8\,(97n^2+538+1174)\ln3 \Bigg]\,.
\end{eqnarray}
For $n=1$ this Gibbs free energy is
given for $T<T_\c$ in Eqs.~(3.14), (3.15) and Table 2 of Ref.~\cite{HD}
where it is denoted by $\tilde{\Gamma}_{-0}(\xi_-,u_0)$.
The logarithmic contributions $O(u_0^3\ln(u_0\xi_\pm))$ 
in Eqs.~(\ref{eq:Gamma(xi+)}) and (\ref{eq:Gamma(xi-)})
correspond to Eq.~(3.15) of Ref.~\cite{HD} and are due
to a specific $d=3$ divergent diagram arising only at four-loop order
\cite{BBMN}.
Its pole term is to be subtracted within the overall subtractions 
mentioned in Section \ref{sec:bare}.

\section{Bare specific heat}
\label{sec:heat}
The specific heat abve (+) and below ($-$) $T_\c$ at $h_0=0$ is determined by
\cite{SD2,Str}
\begin{equation}
\label{eq:def:C}
\bare{C}^\pm = C_B - a_0^2\, \bare{\Gamma}_\pm^{(2,0)}(r_0',u_0)
\end{equation}
where
\begin{equation}
\label{eq:Gamma^(2,0)}
\bare{\Gamma}_\pm^{(2,0)}(r_0',u_0) = \frac{\partial^2}{(\partial r_0')^2}\,
\bare{\cal F}_\pm(r_0',u_0) \,.
\end{equation}
$C_B$ is a noncritical background value and
$a_0$ is a constant defined by $r_0-r_{0\c}=a_0t$ where $t={(T-T_\c)}/T_\c$. 
The result is
\begin{eqnarray}
\label{eq:d_Gamma+}
\bare{\Gamma}_+^{(2,0)}(r_0',u_0) 
               &=& -\frac{n}{16\pi} r_0'^{\,-1/2}
               +\frac{n(n+2)}{2(4\pi)^3} u_0^2 r_0'^{\,-3/2} \Bigg[ n-6 
               -8\ln\frac{3}{4} 
               \nonumber\\
&& \mbox{}     +4\ln \frac{r_0'}{(24u_0)^2}
               \Bigg] + O(u_0^3,u_0^3\ln u_0)
\end{eqnarray}
and
\begin{eqnarray}
\label{eq:d_Gamma-}
\bare{\Gamma}_-^{(2,0)}(r_0',u_0) &=& 
             -\frac{1}{8u_0}
             -\frac{1}{4\pi}(-2r_0')^{-1/2} 
             +\frac{1}{4\pi^2}(n+2)u_0(-2r_0')^{-1}
             \nonumber\\
  && \mbox{} -\frac{u_0^2(-2r_0')^{-3/2}}{384\pi^3} \Bigg[
             16\,(11n+7) -1296c_1 -48(n-1)c_2 
             \nonumber\\
  && \mbox{} +(9n^2+n+17)\pi^2 +36(n-1) \Li{\frac{1}{3}}
             +324 \Li{-\frac{1}{3}} 
             \nonumber\\
  && \mbox{}
             +96(n-1) \left[ \Li{-\frac{1}{2}} -2\Li{-2} \right]
             +18(n-1) (\ln3)^2 \nonumber\\
  && \mbox{}
             +48(n-1) \Big( \ln2 -6\ln3 \Big) \ln2
             -48\,(4n+17)\ln3 \nonumber\\
  && \mbox{} +16\, (31n+95)\ln2 
             +96(n+2)\ln \frac{-2r_0'}{(24u_0)^2} \,\Bigg] \nonumber\\
  && \mbox{} +O(u_0^3,u_0^3\ln u_0) \,.
\end{eqnarray}
Unlike the corresponding two-loop result \cite{Str} for
$\bare{\Gamma}_\pm^{(2,0)}$, the three-loop terms of 
$\bare{\Gamma}_\pm^{(2,0)}$ contain a logarithmic $u_0$-dependence (as long as
they are considered as a function of $r_0'$). This can be absorbed by turning
to the formulation in terms of the correlation lengths $\xi_\pm$
\cite{SD2,SD1}. The relation between $r_0'$ and $\xi_\pm$
is given in Eqs.~(\ref{eq:r0'(xi+)}) and 
(\ref{eq:r0'(xi-)}). 
This leads to the bare perturbative
expression for the specific heat at $h_0=0$ 
that is finite in three dimensions and is free of logarithms in $u_0$,
\begin{eqnarray}
\label{eq:d_Gamma+xi}
\lefteqn{ \bare{\Gamma}_+^{(2,0)}(\xi_+,u_0,3) = \xi_+ \Bigg\{
           -\frac{n}{16\pi} +\frac{n(n+2)}{32\pi^2} u_0\xi_+ } \hspace{0.8cm}
           \nonumber\\
&& \mbox{}
           -\frac{n(n+2)}{1728\pi^3} (u_0\xi_+)^2 \left[ 27n +160 +108
           \ln\frac{3}{4} \right] +O(u_0^3\xi_+^3) \Bigg\}
\end{eqnarray}
and
\begin{equation}
\label{eq:d_Gamma-xi}
\bare{\Gamma}_-^{(2,0)}(\xi_-,u_0,3) = \xi_- \left\{ \sum_{m=0}^3 
a_{-m}^{(2,0)} (u_0\xi_-)^{m-1} +O(u_0^3\xi_-^3) \right\}
\end{equation}
with the coefficients up to two-loop order \cite{Str},
\begin{eqnarray}
\label{eq:a^(2,0)_0}
a_{-0}^{(2,0)} &=& -\frac{1}{8} \,,\\
\label{eq:a^(2,0)_1}
a_{-1}^{(2,0)} &=& -\frac{1}{4\pi} \,,\\
\label{eq:a^(2,0)_2}
a_{-2}^{(2,0)} &=& \frac{3}{8\pi^2}(n+2) \,,
\end{eqnarray}
and the new three-loop coefficient
\begin{eqnarray}
\label{eq:a^(2,0)_3}
a_{-3}^{(2,0)} &=& -\frac{1}{3456\pi^3} \Bigg[
          1188n^2 +11876n +16840 -11664c_1 -432(n-1)c_2
          \nonumber\\
  && \mbox{}
          +9\pi^2 (9n^2+n+17) +324(n-1) \Li{\frac{1}{3}} +2916\Li{-\frac{1}{3}}
          \nonumber\\
  && \mbox{}
          +864(n-1) \left[ \Li{-\frac{1}{2}} -2\Li{-2} \right] 
          +162(n-1) (\ln3)^2 \nonumber\\
  && \mbox{}
          +432(n-1) \Big( \ln2 -6\ln3 \Big) \ln2
          -432\, (4n+17)\ln3 \nonumber\\
  && \mbox{}
          +144\, (31n+95)\ln2\,\Bigg] \,.
\end{eqnarray}
For the case $n=1$, Eqs.~(\ref{eq:d_Gamma-xi})--(\ref{eq:a^(2,0)_3}) agree 
with Eq.~(3.17) and the numerical values in Table 2 of Halfkann and Dohm 
\cite{HD}.

\section{Renormalization and amplitude functions in three dimensions}
\label{sec:amplitude}
In the preceding Sections no renormalizations have been used, except for the
shift of the temperature variable $r_0$ by $\delta r_0$, 
Eq.~(\ref{eq:r0'}), and for the additive subtraction of the Helmholtz free
energy. All other quantities are unrenormalized.
Because of the super-renormalizability of the $\phi^4$ theory for $d<4$,
the shift of
$\delta r_0$ and the additive subtraction are sufficient to make the order
parameter $M_0^2(\xi_-,u_0,3)$ and the vertex function
$\bare{\Gamma}^{(2,0)}_\pm(\xi_\pm,u_0,3)$ finite at infinite cutoff in three
dimensions as long as $\xi_\pm$ is finite.
The resulting bare 
perturbation series in Eqs.~(\ref{eq:M0(xi-)}), (\ref{eq:d_Gamma+xi}) and
(\ref{eq:d_Gamma-xi}), however, are obviously
not applicable near $T_\c$ where $u_0\xi_\pm$ diverges, thus the series
have to be mapped from the critical to the non-critical region.

This mapping is achieved by turning to the renormalized theory as defined
below and by
introducing the renormalization scale $\mu$ which can be varied via
the renormalization-group equation (RGE) \cite{brezin76}. 
The solution of the RGE
implies a decomposition of thermodynamic quantities into a product
of amplitude functions and exponential parts.
This decomposition is most natural and simple within the minimal subtraction
scheme \cite{thooft72} where the exponential parts are determined entirely
from the pure $d=4$ pole terms $\sim\e^{-n}$
of the renormalization constants $Z_r(u,\e)$, $Z_u(u,\e)$,
$Z_\phi(u,\e)$ and $A(u,\e)$ \cite{SD2,SD1}.
As shown previously \cite{SD2,D3,SD1}, the use of these pure pole terms does
not imply the necessity of working near four dimensions and of
using the $\e$-expansion.
This is a nontrivial aspect of the minimal subtraction scheme that
is not widely appreciated in the
field-theoretic literature. Here we make use of the fact that the minimal
subtraction scheme can well
be combined with the concept of a $\phi^4$ theory at fixed dimension $d<4$.

Our $d=3$ approach differs from the $d=3$ approach of
previous work \cite{BBMN,GZ-J,parisi,BNGM,BB,muenster} 
where renormalization conditions were used.
These renormalization conditions are different above and below $T_\c$ whereas
the minimal renormalization applies to both $T>T_\c$ and $T<T_\c$ in the
same form. Within our scheme the minimally
renormalized quantities are introduced in $d$ dimensions as
\begin{eqnarray}
\label{eq:renormalization} 
r = Z_r^{-1}(r_0-r_{0\c})\,, &&\quad
u = \mu^{-\epsilon}A_{d}Z_u^{-1}Z_\phi^2u_0 \,, \quad
\vec\phi = Z_\phi^{-1/2}\vec\phi_0 \,,\\
\label{eq:M^2}
M^2(\xi_-,u,\mu,d) =&& Z_{\phi}^{-1}M_0^2(\xi_-,\mu^\epsilon A_d^{-1} 
                      Z_u Z_\phi^{-2}u,d) \,,\\
\label{eq:Gamma+-}
\Gamma_{\pm}^{(2,0)}(\xi_{\pm},u,\mu,d) =&& Z_r^2\,
\bare{\Gamma}_\pm^{(2,0)}(\xi_{\pm},
       \mu^{\epsilon}Z_uZ_\phi^{-2}A_d^{-1}u,d)\,
 -\, \frac{1}{4}\mu^{-\epsilon}A_dA(u,\epsilon),
\end{eqnarray}
where $\xi_\pm$ are the correlation lengths defined in Refs.~\cite{SD2,SD1}
and
\begin{equation}
\label{eq:A_d}
A_d=\frac{\Gamma(3-d/2)}{2^{d-2} \pi^{d/2} (d-2)}
\end{equation}
is a convenient geometric factor \cite{D3}. 
We recall that the amplitude functions to be defined
in Eqs.~(\ref{eq:def:fphi}) and (\ref{eq:def:F+-}) depend on 
the choice of this geometric factor.
In our calculations,
Eqs.~(\ref{eq:renormalization})--(\ref{eq:A_d}) are used directly at $d=3$.
The analytic form of the renormalization
constants $Z_r(u,\e)$, $Z_u(u,\e)$, $Z_\phi(u,\e)$ and $A(u,\e)$ is given
in Eqs.~(2.13), (2.16)--(2.19) and 
(B1)--(B18) of Ref.~\cite{Lar} for general $n$ up to five-loop order.
As is well known, the minimal renormalization constants in
Eqs.~(\ref{eq:renormalization})--(\ref{eq:Gamma+-}) have the property of
absorbing the remaining ultraviolet divergences for $d\to4$ that were not
absorbed by the mass shift and subtractions mentioned
in Sect.~\ref{sec:bare}. In the context of our $d=3$ theory
this property is, of course, irrelevant. The crucial aspect of the 
renormalization
of Eqs.~(\ref{eq:renormalization})--(\ref{eq:Gamma+-}) at $\e=1$ is that they
provide the mapping and decomposition mentioned above via the integration of
the RGE for the renormalized quantities $M^2$
and $\Gamma^{(2,0)}_\pm$. The critical behavior of these quantities then
evolves from the \em infrared \em divergences of the $Z$-factors $Z_i(u,1)$ as
$u\to u^\star$ \cite{SD1,BB} and from the singular behavior of $A(u,1)$ as 
$u\to u^\star$ \cite{esser}.

Dimensionless amplitude functions $f_\phi$,
$F_+$ and $F_-$ of the renormalized order parameter and specific heat
can be defined at fixed dimension $2<d<4$ according to
\begin{equation}
\label{eq:def:fphi}
f_\phi(\mu\xi_-,u,d) = \xi_-^{d-2} M^2(\xi_-,u,\mu,d)
\end{equation}
and
\begin{equation}
\label{eq:def:F+-}
F_\pm(\mu\xi_\pm,u,d) = -4\mu^\e A_d^{-1}\Gamma_\pm^{(2,0)}(\xi_\pm,u,\mu,d)\,.
\end{equation}
These functions remain finite also in the limit $d\to4$ (at finite $\xi_\pm$)
\cite{SD2,SD1} provided that the appropriate subtractions (including all
additive pole terms for $d<4$)
have been performed
in Sect.~\ref{sec:bare}.
In the application of the solution of the RGE the
amplitude functions appear in the form \cite{SD2} with the non-critical
arguments $\mu\xi_\pm=1$ in three dimensions
\begin{equation}
\label{eq:fphi(3)}
f_\phi(u) \equiv f_\phi(1,u,3)\,, \quad F_\pm(u) \equiv F_\pm(1,u,3)\,.
\end{equation}
These functions should be smooth and well behaved in the entire region
$0\leq u\leq u^\star$ \cite{SD1}. They have the power series
\begin{eqnarray}
\label{eq:series_fphi}
f_\phi(u) &=& \frac{1}{u}\, \sum_{m=0}^\infty c_{\varphi m}^- u^m \,, \\
\label{eq:series_F+}
F_+(u) &=& \sum_{m=0}^\infty c_{Fm}^+ u^m \,, \\
\label{eq:series_F-}
F_-(u) &=& \frac{1}{u}\, \sum_{m=0}^\infty c_{Fm}^- u^m 
\end{eqnarray}
with $n$-dependent coefficients.
These series have a zero radius of convergence but are (presumably) Borel
resummable \cite{SD1,zinn81} (see Sect.~\ref{sec:results}).
Note that the coefficients of these series depend on the choice of the
geometric factor $A_d^{-1}$. Our choice, Eq.~(\ref{eq:A_d}), minimizes the
explicit dimensional dependence of the lowest order coefficients
in Eqs.~(\ref{eq:series_fphi})--(\ref{eq:series_F-})
\cite{SD2,D3,SD1,dohm85} and is expected to improve
the convergence properties of the series. This is of relevance in the context
of Borel resummations based on low-order information.

From Eqs.~(\ref{eq:M0(xi-)}), (\ref{eq:d_Gamma+xi}) and (\ref{eq:d_Gamma-xi})
we obtain the analytic expressions of the coefficients for the order parameter
\begin{eqnarray}
\label{eq:c_phi_0}
c_{\varphi 0}^- &=& \frac{1}{32\pi} \,,\\
\label{eq:c_phi_1}
c_{\varphi 1}^- &=& 0 \,,\\
\label{eq:c_phi_2}
c_{\varphi 2}^- &=& \frac{1}{27\pi}(160-82n) + \frac{2}{\pi}(n-1)\ln3 \,,\\
\label{eq:c_phi_3}
c_{\varphi 3}^- &=& \frac{-1}{1080\pi} \Bigg\{\,
                2500n^2 +65104n +29056 +8640\, (5n+22) \zeta(3)
                +58320c_1 \nonumber\\
   &&\mbox{}    +2160(n-1) \left[ 4\Li{-\frac{1}{2}} -c_2 \right]
                -15\pi^2 (19n^2+643n+499) 
                \nonumber\\
   &&\mbox{}    -180\,(64n^2+640n+457) \Li{-\frac{1}{3}} +11340(n-1) 
                \Li{\frac{1}{3}} \nonumber\\
   &&\mbox{}    -17280(n-1) \Li{-2}
                +5670(n-1) (\ln3)^2 +4320(n-1) (\ln2)^2
                \nonumber\\
   &&\mbox{}    -25920(n-1) (\ln2)(\ln3) 
                -80\, (194n^2+1616n-1675)\ln3 \nonumber\\
   &&\mbox{}    +16\, (860n^2+8357n-7867) \ln2\,\Bigg\}\,,
\end{eqnarray}
for the specific heat above $T_\c$,
\begin{eqnarray}
\label{eq:c_F0^+}
c_{F0}^+ &=& -n \,,\\
\label{eq:c_F1^+}
c_{F1}^+ &=& -2n(n+2) \,,\\
\label{eq:c_F2^+}
c_{F2}^+ &=& -4n(n+2) \left[n-\frac{7}{27}+4\ln\frac{4}{3} \right] \,,
\end{eqnarray}
and for the specific heat below $T_\c$,
\begin{eqnarray}
\label{eq:c_F0^-}
c_{F0}^- &=& \frac{1}{2} \,,\\
\label{eq:c_F1^-}
c_{F1}^- &=& -4\,, \\
\label{eq:c_F2^-}
c_{F2}^- &=& 8(10-n) \,,\\
\label{eq:c_F3^-}
c_{F3}^- &=&   -\frac{1}{27} (1080n^2 +3464n +31120)
               -128\, (5n+22)\zeta(3) -864c_1 \nonumber\\
   && \mbox{}  -32(n-1)c_2 +\frac{\pi^2}{3} (18n^2 +2n +34)
               +24(n-1) \Li{\frac{1}{3}}  \nonumber\\
   && \mbox{}  +216 \Li{-\frac{1}{3}} 
               +64(n-1) \Li{-\frac{1}{2}}
               -128(n-1) \Li{-2} 
               \nonumber\\
   && \mbox{}  +12(n-1) (\ln3)^2 
               +32(n-1) (\ln2)^2 -192(n-1) (\ln2)(\ln3) 
               \nonumber\\
   && \mbox{}  -32\, (4n+17)\ln3 +\frac{32}{3} (31n+95) \ln2 \,.
\end{eqnarray}
The terms up to $O(u)$ are the previous two-loop results \cite{SD2,Str}.
In Table \ref{table:coefficients} we give the numerical values of the
coefficients $c_{\varphi m}^-$, $c_{Fm}^+$ and $c_{Fm}^-$
up to three-loop order for $n=1$, 2, 3 as obtained from
Eqs.~(\ref{eq:c_phi_0})--(\ref{eq:c_F3^-}).

\begin{table}
\caption{Coefficients $c^-_{\varphi m}$ of $f_\phi(u)$ according to 
Eqs.~(\ref{eq:series_fphi}), (\ref{eq:c_phi_0})--(\ref{eq:c_phi_3}),
and $c_{Fm}^\pm$ of $F_+(u)$ and $F_-(u)$ 
according to Eqs.~(\ref{eq:series_F+}), (\ref{eq:c_F0^+})--(\ref{eq:c_F2^+})
and (\ref{eq:series_F-}), (\ref{eq:c_F0^-})--(\ref{eq:c_F3^-}),
respectively, for $n=1$, 2, 3 up to three-loop order. 
For $c_{\varphi m}^-$ and $c_{Fm}^-$, $m$ refers to $u^{m-1}$ corresponding
to $m$-loop order
whereas for $c_{Fm}^+$, $m$ refers to $u^m$ corresponding to $(m+1)$-loop 
order.}
\begin{tabular}{ccrrr}
 & $m$ & $c^-_{\varphi m}$ & $c_{Fm}^+$ & $c_{Fm}^-$   \\
\hline
$n=1$  & 0 & $(32\pi)^{-1}$& -1            & 0.5 \\
       & 1 & 0             & -6            & -4 \\
       & 2 & 0.919561893   & -22.6976284   &  72 \\
       & 3 & -80.8015258   &               & -5189.75474 \\
& \multicolumn{3}{c}{ } &\\
$n=2$  & 0 & $(32\pi)^{-1}$& -2            & 0.5 \\
       & 1 & 0             & -16           & -4 \\
       & 2 & 0.652241285   & -92.5270090   & 64 \\
       & 3 & -83.0064428   &               & -5918.07320 \\
& \multicolumn{3}{c}{ } &\\
$n=3$  & 0 & $(32\pi)^{-1}$& -3            & 0.5 \\
       & 1 & 0             & -30           & -4 \\
       & 2 & 0.384920676   & -233.488142   & 56 \\
       & 3 & -82.6969869   &               & -6607.95641 \\ \hline
\end{tabular}
\label{table:coefficients}
\end{table}

For $n=1$ the numerical values of the
coefficients $c_{\varphi m}^-$ given previously \cite{HD} agree
with ours up to eight digits. The difference in the ninth digit is due to a
numerical inaccuracy of the $Z$-factors in three dimensions that were available
previously only in numerical form \cite{KSD}.
(A similar comments applies to the numerical values of the series for $n=1$
\cite{HD} mentioned in the preceding Sections.)
In the present work we have used
the $Z$-factors in the analytic form for general $n$ as given in 
Ref.~\cite{Lar}.

Numerical results of a six-loop calculation (above $T_\c$ for $n=1$, 2, 3) and
of a five-loop calculation (below $T_\c$ for $n=1$) 
of the diagrams contributing to $F_\pm$ in three dimensions
have been presented by Baker et al.\ \cite{BNGM},
by Bagnuls and Bervillier \cite{BB} and by Bagnuls et al.\ 
\cite{BBMN}, respectively.
On this basis the coefficients $c_{Fm}^+$ for $n=1$, 2, 3 and $c_{Fm}^-$
for $n=1$ were calculated in numerical form
by Krause et al.\ \cite{KSD} and by Halfkann and Dohm \cite{HD}
who used a two-loop approximation for the additive renormalization
constant $A(u,\e)$. The corrected coefficients for $c_{Fm}^+$ for $n=1$, 2, 3 
and $c_{Fm}^-$ for $n=1$ using $A(u,\e)$ in five-loop order have been presented
recently in numerical form \cite{Lar}.
Our present results, Eqs.~(\ref{eq:c_F0^+})--(\ref{eq:c_F3^-}), 
provide the analytic
form of these coefficients up to three-loop order for general $n$. The 
three-loop coefficient $c_{F2}^+$ 
has been calculated independently by Burnett \cite{B}.

\section{Results and discussion}
\label{sec:results}
Within the minimally renormalized $\phi^4$ field theory in three dimensions
we have derived the three-loop contributions to the amplitude functions
$f_\phi(u)$ of the square of the order parameter, Eqs.~(\ref{eq:series_fphi}),
(\ref{eq:c_phi_3}), and
to the amplitude functions $F_+(u)$ and $F_-(u)$ of the specific heat above
and below $T_\c$, Eqs.~(\ref{eq:series_F+}), (\ref{eq:c_F2^+})
and (\ref{eq:series_F-}), (\ref{eq:c_F3^-}) for general $n$.
Similar to the previous results below $T_\c$ for $n=1$ \cite{BBMN,HD,Lar},
the coefficients $c_{\varphi m}^-$ and $c_{Fm}^-$ for $n=2$ and $n=3$ given
in Table \ref{table:coefficients} have alternating signs and increase
considerably in magnitude.
Clearly a resummation of these series is necessary in order to obtain 
quantitatively reliable results. For a description of the method of Borel
resummation we refer to Refs.~\cite{SD1,zinn81} and, in the present context,
to Ref.~\cite{Lar}.

While the previous two-loop results \cite{Str} did not yet provide sufficient
information for a controlled resummation procedure and thus did not yet lead
to an error estimate, the new information on $c_{\varphi3}^-$ and $c_{F3}^-$
presented here makes possible to perform the Borel resummation for $n>1$
below $T_\c$ with reasonably small error bars. The reliability of this Borel
resummation based on three-loop results (with four low-order coefficients,
see Table \ref{table:coefficients})
has been demonstrated in detail in a related paper \cite{Lar}. We refer to
this paper for a description of the method of determining the error bars.
For the case $n=2$ (corresponding to the superfluid transition of $^4$He)
the resummation result is
\begin{equation}
\label{eq:F-star}
u^\star F_-(u^\star) = 0.384\, \pm\, 0.025
\end{equation}
at the fixed point $u^\star=0.0362$ \cite{Lar}. 
In addition, using the new coefficient $c_{\varphi3}^-$ for $n=2$ and
$n=3$, a Borel
resummation of the fixed point value of the amplitude function 
$f_\phi(u^\star)$ has been performed by M\"onnigmann \cite{Moennigmann}.
(For this case the Borel resummation parameters $b$ and $\alpha$
\cite{SD1,zinn81} are $b=3.5+n/2=4.5$ and $-0.43\leq\alpha\leq0.47$ for $n=2$,
and $b=5$, $-0.62\leq\alpha\leq0.71$ for $n=3$.)
The result is
\begin{eqnarray}
\label{eq:fphi_star_n=2}
u^\star f_\phi(u^\star) &=& 0.010099\, \pm\, 0.000084\,, \quad n=2\,, \\
\label{eq:fphi_star_n=3}
u^\star f_\phi(u^\star) &=& 0.00997\, \pm\, 0.00011\,, \quad n=3
\end{eqnarray}
at the fixed point \cite{Moennigmann}. 
Thus the deviation from the zero-loop term 
$c_{\varphi 0}^-=(32\pi)^{-1}\approx0.009947$ is very small,
confirming the previous expectation \cite{SD2}.

The amplitude functions are of physical relevance not only at the fixed point
but also in a finite interval $0<u<u^\star$ corresponding to a finite distance
from criticality where non-asymptotic corrections become non-negligible
\cite{D2}.
Since the Borel resummed amplitude functions $F_-(u)$ and $f_\phi(u)$ are
smooth and slowly varying functions for $n=1$ (see Figs.~1 and 4 of
Ref.~\cite{HD}) the same property should hold also for $n=2$ and 3.
Empirical evidence in support 
of this fact comes from analyses of experimental data of $^4$He in the
non-asymptotic region \cite{dohm84,D2,schloms87}. Therefore, within the
accuracy of the present results, these functions can be approximated 
by the simple representations \cite{HD,KSD}
\begin{equation}
\label{eq:rep_F}
F_-(u) = (2u)^{-1} -4(1+d_Fu) 
\end{equation}
and
\begin{equation}
\label{eq:rep_fphi}
f_\phi(u) =A_3 (8u)^{-1} (1+d_\varphi u) \,,
\end{equation}
with $A_3=(4\pi)^{-1}$. Here the lowest-order terms are chosen to coincide
with the one-loop results. The coefficients $d_F$ and $d_\varphi$ are
determined by requiring that Eqs.~(\ref{eq:rep_F}) and (\ref{eq:rep_fphi})
agree with the Borel-resummed fixed point values of Eqs.~(\ref{eq:F-star})
and (\ref{eq:fphi_star_n=2}). This implies
\begin{equation}
\label{eq:d_F}
d_F=-5.49\,, \quad d_\varphi=0.422
\end{equation}
for $n=2$. In view of the present error bars, a more refined representation
of $F_-(u)$ and $f_\phi(u)$ does not seem to be warranted until four-loop
(Borel-resummed) results with smaller error bars become available.

\begin{figure}[h]
\hspace{1.5cm}\psfig{figure=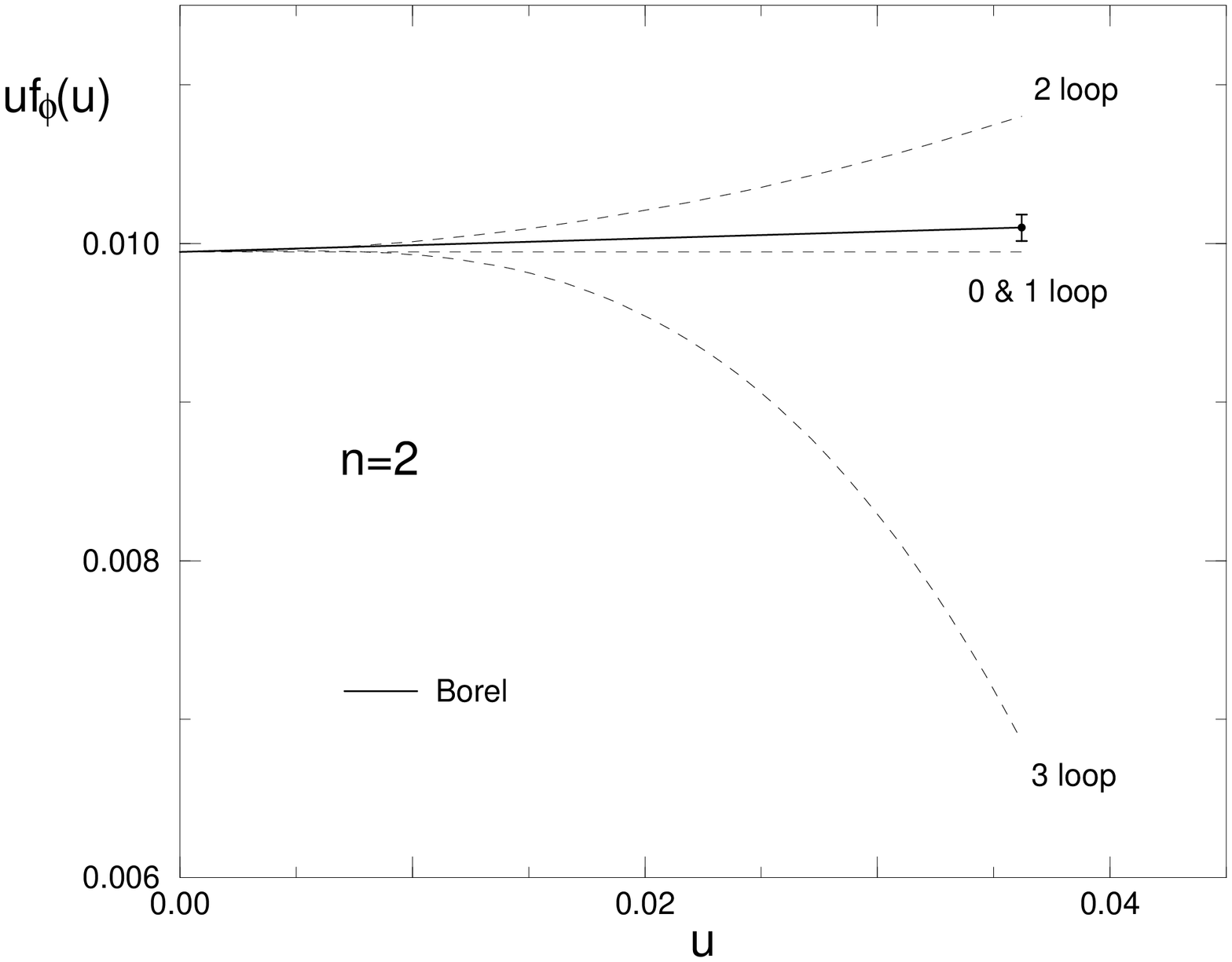,angle=-90,width=9cm}
\caption{
Partial sums $f^{(M)}_\phi(u) = \frac{1}{u}\sum_{m=0}^M c_{\varphi m}^-u^m$
of the amplitude function $f_\phi(u)\equiv f_{\phi}(1,u,3)$,
Eq.~(\ref{eq:series_fphi}), 
for the square of the order parameter in three dimensions for $n=2$ multiplied 
by $u$, as a function of the
renormalized coupling $u$ from $M=0$ (zero-loop order) to $M=3$ (three-loop
order) (dashed lines). 
The solid curve is Eq.~(\ref{eq:rep_fphi}) with $d_\varphi=0.422$
representing the Borel summation result.
The error bar indicates the error of the Borel resummed fixed point
value $f_\phi(u^\star)$ at $u^\star=0.0362$, Eq.~(\ref{eq:fphi_star_n=2}).
}
\label{figure:M}
\end{figure}

\begin{figure}[h]
\hspace{1.5cm}\psfig{figure=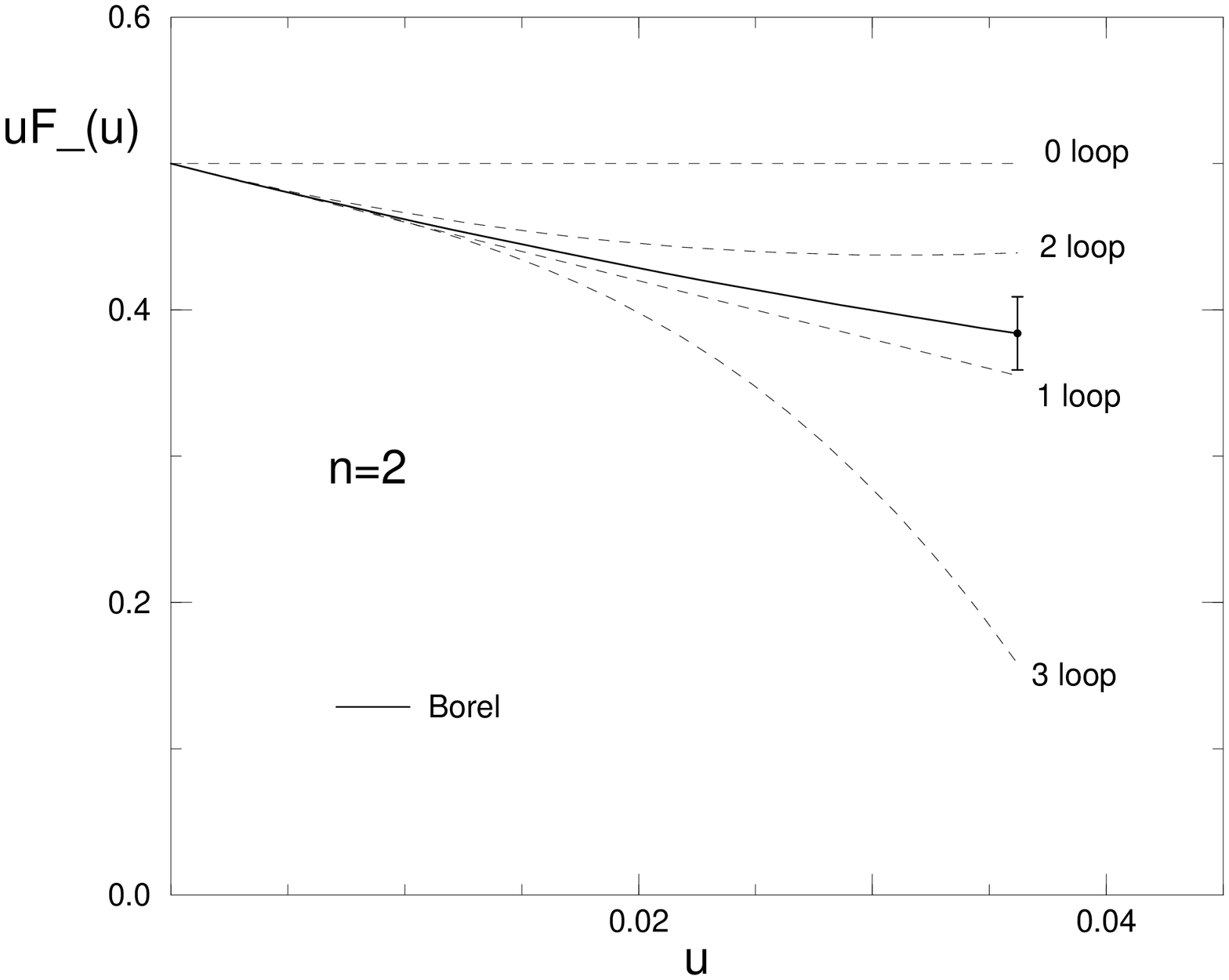,angle=-90,width=9cm}
\caption{
Partial sums $F^{(M)}_-(u) = \frac{1}{u}\sum_{m=0}^M c_{Fm}^-u^m$
of the amplitude function $F_-(u)\equiv F_-(1,u,3)$,
Eq.~(\ref{eq:series_F-}), 
for the specific heat below $T_\c$ in three dimensions for $n=2$ multiplied 
by $u$, as a function of the
renormalized coupling $u$ from $M=0$ (zero-loop order) to $M=3$ (three-loop
order) (dashed lines). 
The solid curve is Eq.~(\ref{eq:rep_F}) with $d_F=-5.49$ representing 
the Borel summation result.
The error bar indicates the error of the Borel resummed fixed point
value $F_-(u^\star)$ at $u^\star=0.0362$, Eq.~(\ref{eq:F-star}).
}
\label{figure:F}
\end{figure}

In Figs.~\ref{figure:M} and \ref{figure:F} we have plotted
$f_\phi(u)$ and $F_-(u)$ for $n=2$ according to the
representations in Eqs.~(\ref{eq:rep_F}) and (\ref{eq:rep_fphi})
(solid lines), together with the zero-,
one-, two- and three-loop approximations in the range $0\leq u\leq u^\star$ 
(dashed lines).
The amplitude functions $f_\phi(u)$ and $F_\pm(u)$ enable us to calculate and
predict
(rather than fit) the various ratios of amplitudes of leading and subleading
terms of the critical temperature dependence of the order parameter and the
specific heat \cite{SD2}. These ratios are universal quantities \cite{PHA}.

As a first application we consider the specific heat which plays an
important role in testing the renormalization-group prediction of
critical-point universality along the $\lambda$-line of $^4$He
\cite{D1,LDID}.
Combining the three-loop result for $F_-(u^\star)$, Eq.~(\ref{eq:F-star}),
with the recent \cite{Lar} five-loop result for the additive renormalization
constant $A(u,\e)$ leads to the
universal amplitude ratio \cite{Lar}
\begin{equation}
\label{eq:A+/A-}
\frac{A^+}{A^-} = 1.056 \pm 0.004
\end{equation}
for the asymptotic specific heat in three dimensions. This
is in excellent agreement
with the high-precision experimental result $A^+/A^-=1.054\pm0.001$ for
$^4$He near the superfluid transition
obtained from a recent experiment in space \cite{lipa96}. Our result,
Eq.~(\ref{eq:A+/A-}), is a
significant improvement over the previous prediction \cite{bervillier76}
$A^+/A^-=1.0294\pm 0.0134$ obtained from the $\e=4-d$ expansion up to
$O(\e^2)$ \cite{see}.

As a second application we employ our Borel-resummed
three-loop result for $f_\phi$
to calculate the universal combination of amplitudes \cite{PHA}
\begin{equation}
\label{eq:R_C}
R_C \equiv
\frac{A_\chi A^+}{A_M^2} = \frac{(2\nu P^\star_+)^{2-2\beta}}
{(3/2 - 2 \nu P^\star_+)^{2\beta}} \cdot
\frac{4\nu B^\star + \alpha F_+^\star}{16\pi}\:
\left[f_\phi^\star f^{(2)\star}\right]^{-1}
\end{equation}
where $A_\chi$ and $A_M$ are the leading amplitudes of the susceptibility above
$T_\c$ and of the order parameter below $T_\c$, respectively \cite{PHA}.
The structure of the right-handed side of Eq.~(\ref{eq:R_C}) follows from
previous work \cite{SD2,KSD}.
Here $f_\phi^\star=f_\phi(u^\star)$ is determined by 
Eqs.~(\ref{eq:fphi_star_n=2}) and (\ref{eq:fphi_star_n=3}).
The quantities $P^\star=P(u^\star)$, $F_+^\star=F_+(u^\star)$,
$B^\star=B(u^\star)$ and $f^{(2)\star}=f^{(2)}(u^\star)$ have been calculated
in Refs.~\cite{Lar} and \cite{KSD}, respectively. The values for the critical
exponents $\alpha$ and $\nu$ are given in Ref.~\cite{Lar}, the value of $\beta$
is taken from Ref.~\cite{LZJ}.
The resulting values for $R_C$ in three dimensions are
\begin{eqnarray}
\label{eq:R_C_n=2}
R_C &=& 0.123\,\pm\,0.003\,, \quad n=2\,, \\
\label{eq:R_C_n=3}
%
R_C &=& 0.189\,\pm\,0.009\,, \quad n=3\,.
\end{eqnarray}
These results are significantly more accurate than the $\e$-expansion values
\cite{AH} up to $O(\e^2)$ whose extrapolation to $\e=1$ suffers from
considerable ambiguities \cite{AH}.
The numerical
estimate $R_C=0.165$ of high-temperature series expansions for $n=3$
\cite{Milo} is not far from our value, Eq.~(\ref{eq:R_C_n=3}).
Additional numerical and experimental studies would be desirable.

Further applications of our results will be given elsewhere.

%
%
%
%
%

\section*{Acknowledgements}
We gratefully acknowledge support by Deutsches Zentrum f\"ur Luft- und 
Raumfahrt (DLR, previously DARA) under grant 
number 50 WM 9669 as well as by NASA under contract number 960838.

\appendix
\section{Three-loop vacuum diagrams}
\label{app:diagrams}
In this Appendix we present 
some details of the calculation (at infinite cutoff)
of the three-loop vacuum diagrams shown in
Fig.~\ref{fig:vacuum} in three dimensions.
There are six different topologies of vacuum diagrams (denoted as
(A), (B), \ldots (F) in Ref.~\cite{Raj}) 
whose integral expressions are given by
\mathindent-1cm
\begin{eqnarray}
\label{eq:diagram_A}
\Dia{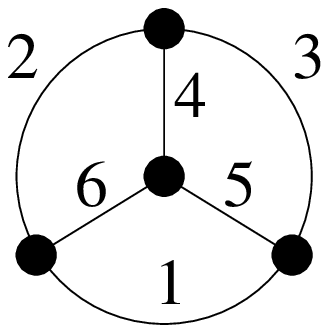}{3.9}{10.3}\: &=& 
\int_{\p^{}_1} \!\int_{\p^{}_2} \!\int_{\p^{}_3}
\textstyle\frac{1}{(m_1^2+p_1^2) (m_2^2+|\p^{}_1-\p^{}_2|^2) (m_3^2+p_2^2)
(m_4^2+|\p^{}_2-\p^{}_3|^2) (m_5^2+p_3^2) (m_6^2+|\p^{}_3-\p^{}_1|^2)} \,, 
\nonumber\\ &&\\
\label{eq:diagram_B}
\Dia{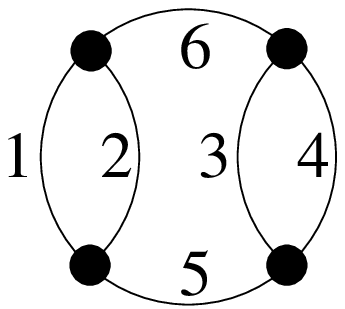}{4}{11}\: &=& 
\int_{\p^{}_1} \!\int_{\p^{}_2} \!\int_{\p^{}_3}
\textstyle\frac{1}{(m_1^2+p_1^2) (m_2^2+|\p^{}_3-\p^{}_1|^2) 
(m_3^2+|\p^{}_3-\p^{}_2|^2) 
(m_4^2+p_2^2) (m_5^2+p_3^2) (m_6^2+p_3^2)}\,, \\
\label{eq:diagram_C}
\Dia{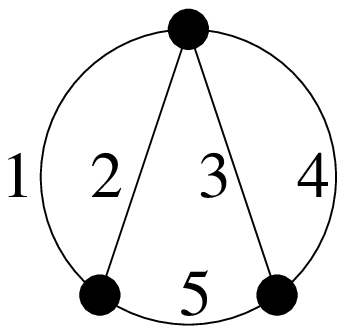}{3.7}{11}\: &=&  
\int_{\p^{}_1} \!\int_{\p^{}_2} \!\int_{\p^{}_3}
\textstyle\frac{1}{(m_1^2+p_1^2) (m_2^2+|\p^{}_3-\p^{}_1|^2) 
(m_3^2+|\p^{}_3-\p^{}_2|^2) (m_4^2+p_2^2) (m_5^2+p_3^2)} \,,\\
\label{eq:diagram_D}
\hspace*{1cm}
\Dia{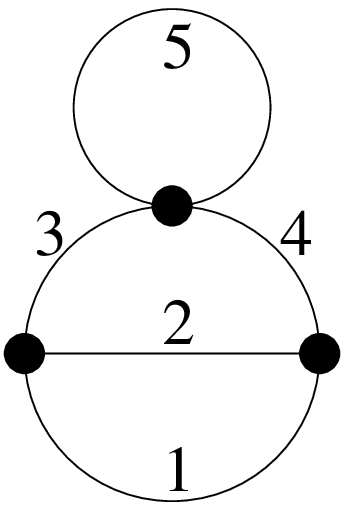}{4.0}{11.3}\: &=&  
\int_{\p^{}_1} \!\int_{\p^{}_2} \!\int_{\p^{}_3}
\textstyle\frac{1}{(m_1^2+p_1^2) (m_2^2+p_2^2) (m_3^2+|\p^{}_1+\p^{}_2|^2) 
(m_4^2+|\p^{}_1+\p^{}_2|^2) (m_5^2+p_3^2)} \,,\\
\label{eq:diagram_E}
\hspace*{1cm}
\Dia{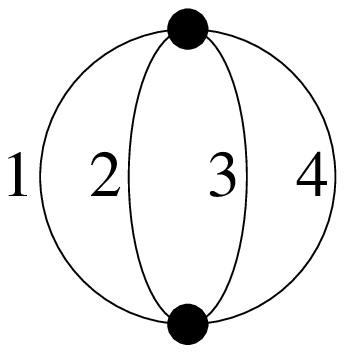}{5}{11.5} \: &=&  
\int_{\p^{}_1} \!\int_{\p^{}_2} \!\int_{\p^{}_3}
\textstyle\frac{1}{(m_1^2+p_1^2) (m_2^2+p_2^2) (m_3^2+p_3^2) 
(m_4^2+|\p^{}_1+\p^{}_2+\p^{}_3|^2)} \,, \\
\label{eq:diagram_F}
\hspace*{1cm}
\Dia{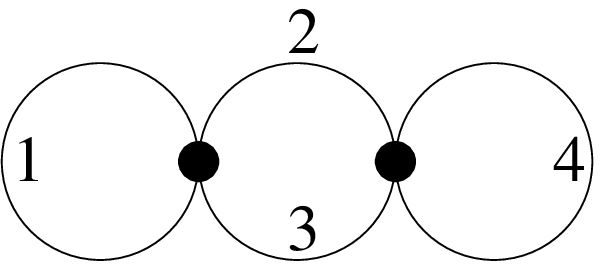}{2.2}{20}\: &=&  
\int_{\p^{}_1} \!\int_{\p^{}_2} \!\int_{\p^{}_3}
\textstyle\frac{1}{(m_1^2+p_1^2) (m_2^2+p_2^2) (m_3^2+p_2^2) (m_4^2+p_3^2)} \,.
\end{eqnarray}
In our case, the masses $m_i^2$ are given by $\rol$ or $\rot$ of
Eq.~(\ref{eq:rol}).
Three of the diagrams in Fig.\ 1 are infrared divergent in the
limit $\rot\to0$ in three dimensions: type (B) with $m_1^2=m_4^2=\rol$ and 
$m_2^2=m_3^2=m_5^2=m_6^2=\rot$, type (D) with $m_1^2=m_2^2=m_5^2=\rol$ and
$m_3^2=m_4^2=\rot$ and type (F) with $m_1^2=m_4^2=\rol$ and
$m_2^2=m_3^2=\rot$. 
The leading $\rot$ dependence of these diagrams appears in the terms of
$O(\bar{w}^{-1/2})$ in Eqs.~(\ref{eq:F300}), (\ref{eq:F310-}) and
(\ref{eq:F320-}). All other diagrams in Fig.\ 1 are free of
infrared divergences in the limit $\rot\to0$ at $d=3$.

Most difficult is the evaluation of the ``Mercedes'' diagram (A), 
Eq.~(\ref{eq:diagram_A}). It has been calculated by Rajantie \cite{Raj}
in three
dimensions only for two special cases (i) where one of the masses $m_i^2$ 
vanishes, or (ii) where all the masses are equal. The case (ii) corresponds
to our first Mercedes diagram in Fig.~\ref{fig:vacuum} with $m_i^2=\rol$ and
is given by Eq.~(55) of Ref.~\cite{Raj}. The last two Mercedes diagrams in
Fig.~\ref{fig:vacuum}, however, contain the two different (longitudinal and
transverse) masses $\rol$ and $\rot$.
This case is considerably more difficult than the cases (i) and (ii) mentioned
above. We have been able to calculate these diagrams for small
$\bar{w}=\rot/\rol>0$ in three dimensions as
\mathindent1cm
\begin{eqnarray}
\label{eq:Merc1}
\Dia{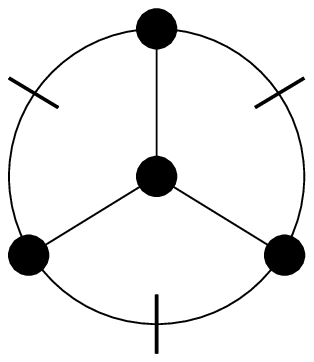}{5.5}{10.3}\: &=& \frac{\rol^{-3/2}}{(4\pi)^3}\,
\Bigg[\: \frac{3}{2} \Li{-2} + \frac{3}{2} (\ln 3)(\ln 2) + \frac{\pi^2}{8}
+ 3\bar{w}^{1/2} \ln \frac{3}{4} \nonumber\\
&&\phantom{\frac{\rol^{3/2}}{(4\pi)^3}}
+ \frac{3}{4}\,\bar{w} \left( 3\Li{-2} + 3(\ln 3)(\ln 2) + 4\ln2
+ \frac{\pi^2}{4} \right) \nonumber\\
&&\phantom{\frac{\rol^{3/2}}{(4\pi)^3}}
+O(\bar{w}^{3/2}, \bar{w}^{3/2}\ln\bar{w}) \Bigg] \,, \\
\label{eq:Merc2}
\Dia{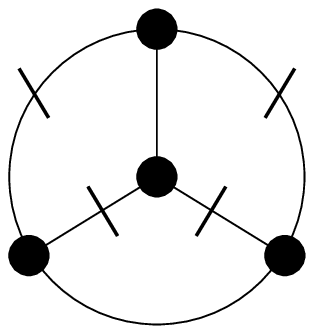}{3.9}{10.3}\: &=& \frac{\rol^{-3/2}}{(4\pi)^3}\,
\Bigg[\: c_2 +2\bar{w}^{1/2} (\ln \bar{w} +6\ln 2 -2) +\bar{w}(c_2-6-4\ln2)
\nonumber\\
&&\phantom{\frac{\rol^{3/2}}{(4\pi)^3}}
+O(\bar{w}^{3/2}, \bar{w}^{3/2}\ln\bar{w}) \Bigg] \,,
\end{eqnarray}
where $c_2$ given by Eq.~(\ref{eq:c2}).

For finite $m_i^2$ the diagrams (B)--(D) and (F) are finite in three 
dimensions at infinite cutoff whereas (E), Eq.~(\ref{eq:diagram_E}), has a
single pole $\sim (d-3)^{-1}$. Analytic results for the diagrams (B)--(D) and
(F), Eqs.~(\ref{eq:diagram_B})--(\ref{eq:diagram_D}) and (\ref{eq:diagram_F}),
are given in Eqs.~(A.24), (A.25) and (A.27) of Ref.~\cite{Raj} and in Eq.~(5)
of the Erratum \cite{Raj}.

Regarding the $d=3$ divergent diagram (E), the $\overline{\mathrm{MS}}$ scheme 
was used in Ref.~\cite{Raj} which differs from our approach. 
Within our approach the result for diagram (E) near $d=3$ is represented as
\begin{eqnarray}
\label{eq:result_E}
u_0^2\; \Dia{3-loopE_numbered}{5}{11.5}\; = \frac{u_0^{-1+d/\e}}{(4\pi)^3}
\sum_{i=1}^4 m_i
\Bigg[ && -\frac{1/2}{\e-1} -2 +\frac{3}{4}\gamma 
-\frac{1}{4}\ln \Big[ 16\pi^3 \Big] \nonumber\\
&& \mbox{}
+ \frac{1}{2} \ln \frac{m_i}{u_0}
+ \ln \frac{\sum_i m_i}{u_0} +O(\e-1) \Bigg]
\end{eqnarray}
with $\gamma\equiv C_{\mathrm Euler}$ being Euler's constant and $\e=4-d$.
The pole term $\sim (\e-1)^{-1}$ is cancelled by a term of order $O(u_0^2)$
that arises from performing the mass shift $r_0=r_0'+\delta r_0$ in the
one-loop contribution of the Helmholtz free energy.
Eq.~(\ref{eq:result_E}) corresponds to Eqs.~(21) and (A.26) of Ref.~\cite{Raj}
and Eq.~(1) of the Erratum \cite{Raj}. Note that in
our Eq.~(\ref{eq:result_E}) there is no additional parameter corresponding
to the $\overline{\mathrm{MS}}$ scale parameter $\bar{\mu}$ of Ref.~\cite{Raj}.

\section{Correlation lengths in three-loop order}
\label{app:correlation}
In this Appendix we derive Eqs.~(\ref{eq:r0'(xi+)}) and (\ref{eq:r0'(xi-)}).
Above $T_\c$, the correlation length is defined via
\begin{equation}
\label{eq:correlation}
\xi_+^2 = \left. \bare{\chi}_+(0)\,
\partial\bare{\chi}_+(q)^{-1}/\partial q^2 \right|_{q=0}
\end{equation}
where $\bare{\chi}_+(q)^{-1}=\Gamma_0^{(0,2)}(q,r_0,u_0)$ is the
inverse susceptibility at finite wavenumber $q$. The two-point vertex function 
$\Gamma_0^{(0,2)}(q,r_0,u_0)$ is given by
\begin{eqnarray}
\label{eq:Gamma^(0,2)}
\Gamma_0^{(0,2)}(q,r_0,u_0) = r_0+q^2 -\Sigma_0(q,r_0,u_0) 
\end{eqnarray}
where the self-energy 
\begin{equation}
\label{eq:sigma}
\Sigma_0(q,r_0,u_0) = \sum_{m=1}^\infty (-u_0)^m \Sigma_0^{(m)}(q,r_0)
\end{equation}
is the sum of all one-particle 
irreducible $m$-loop diagrams with two (amputated)
external legs. In Eqs.~(A.1)--(A.5) of Ref.~\cite{Str} the functions
$\Gamma_0^{(0,2)}(q,r_0,u_0)$, Eq.~(\ref{eq:Gamma^(0,2)}),
and $\xi_+(r_0,u_0)$, Eq.~(\ref{eq:correlation}),
are given in their two-loop approximation.
The diagrams of the three-loop contribution to Eq.~(\ref{eq:sigma})
are given by
\begin{eqnarray}
\label{eq:sigma^(3)}
\Sigma_0^{(3)}(q,r_0) &=& 64(n+2)^3\;  \Dia{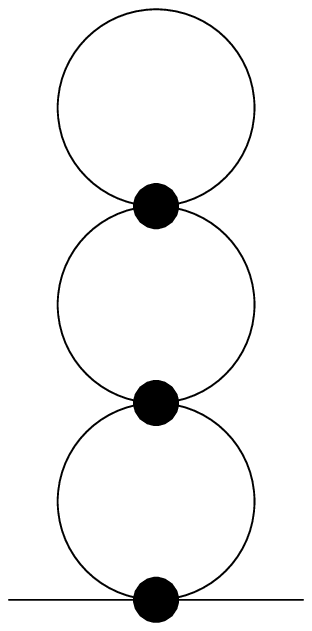}{0}{10}\;
                    +64(n+2)^3\!\!\! \Dia{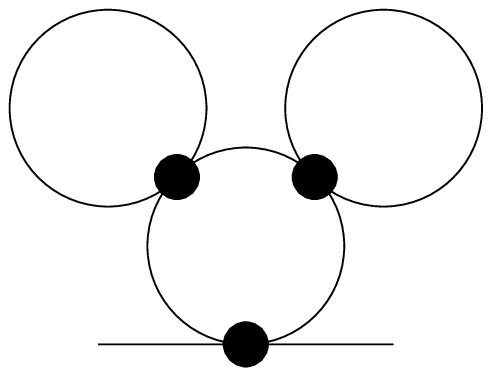}{0}{16}
                    +128(n+2)^2\;    \Dia{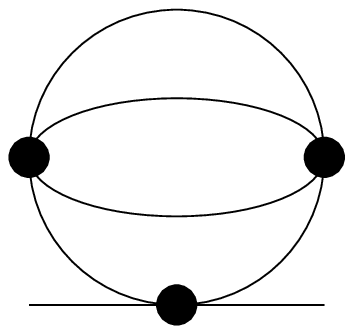}{-.5}{11.5} \nonumber\\
         && \mbox{} +384(n+2)^2\;    \Dia{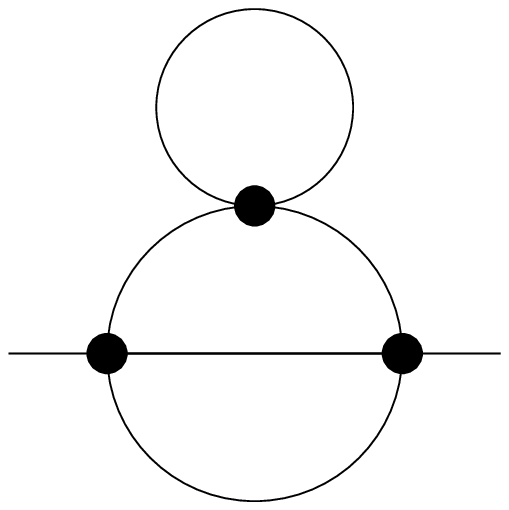}{3.8}{16}\;
                    +128(n+2)(n+8)\; \Dia{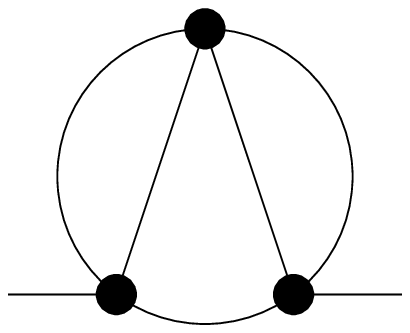}{0}{13}\;.
\end{eqnarray}
For $q=0$ the lines denote the
standard propagator $(r_0+p^2)^{-1}$ above $T_\c$.
The derivative of $\Sigma_0^{(3)}(q,r_0)$ with respect to $q^2$ yields
\begin{eqnarray}
\label{eq:d_sigma^(3)_dq^2}
\lefteqn{\left.\frac{\partial}{\partial q^2} 
\Sigma_0^{(3)}(q,r_0)\right|_{q=0}=} \nonumber\\
&&\mbox{} -384(n+2)^2 \Bigg[\, \frac{\e}{d}\frac{\partial}{\partial m_1^2}\; 
\Dia{3-loopD_numbered}{4}{11.3}\; 
+\frac{2r_0}{d}\, \frac{\partial^2}{(\partial m_1^2)^2}\;
\Dia{3-loopD_numbered}{4}{11.3}\: \Bigg]_{m_i^2=r_0} \nonumber\\
&&\mbox{} -128(n+2)(n+8) \Bigg[\,\frac{\e}{d}\frac{\partial}{\partial m_5^2}\; 
\Dia{3-loopC_numbered}{3.7}{11}\; 
+\frac{2r_0}{d}\, \frac{\partial^2}{(\partial m_5^2)^2}\;
\Dia{3-loopC_numbered}{3.7}{11}\: \Bigg]_{m_i^2=r_0}.
\end{eqnarray}
All diagrams appearing in Eqs.~(\ref{eq:sigma^(3)}) and
(\ref{eq:d_sigma^(3)_dq^2}) can be calculated (at $d=3$ and $q=0$)
using Appendix \ref{app:diagrams} with $m_i^2\equiv r_0$.

Following Ref.~\cite{Str} we 
invert $\xi_+(r_0,u_0)$, Eq.~(\ref{eq:correlation}),
and get the function $r_0(\xi_+,u_0)$.
Subtracting $\delta r_0=r_0-r_0'$, Eq.~(\ref{eq:r0'}),
from $r_0(\xi_+,u_0)$ and letting $d\to3$
we obtain Eq.~(\ref{eq:r0'(xi+)}). Below $T_\c$ we get $r_0'(\xi_-,u_0)$,
Eq.~(\ref{eq:r0'(xi-)}), at $d=3$ as described in Ref.~\cite{Str}.

%
%
%

%
%
%
%


\begin{thebibliography}{99}
\bibitem{see}
See also Appendix C of Ref.~\protect\cite{BBMN} where the $\e$-expansion
result for $A^+/A^-$ is given up to $O(\e^2)$ for general $n$ without
resummation. Setting $\e=1$ in this expression would yield $A^+/A^-=0.7815$
for $n=2$.
\end{thebibliography}
\end{document}